\documentclass[aps,superscriptaddress,amsmath,amssymb,twocolumn,10pt,floatfix,pra]{revtex4}
\usepackage{graphicx}
\usepackage{epsfig}
\usepackage{amssymb,amsfonts,amsmath}
\begin{document}

\title{Neutral skyrmion configurations in the low-energy effective theory of
spinor condensate ferromagnets}
\author{R. W. Cherng}
\author{E. Demler}
\affiliation{Physics Department, Harvard University, Cambridge, MA
02138}
\date{\today}
\begin{abstract}
We study the low-energy effective theory of spinor condensate
ferromagnets for the superfluid velocity and magnetization 
degrees of freedom.  This effective theory describes the competition between
spin stiffness and a long-ranged interaction between skyrmions,
topological objects familiar from the theory of ordinary ferromagnets.
We find exact solutions to the non-linear equations
of motion describing neutral configurations of skyrmions and anti-skyrmions.
These analytical solutions provide a simple physical picture for the origin
of crystalline magnetic order in spinor condensate ferromagnets with dipolar
interactions.
We also point out the connections to effective theories for quantum Hall 
ferromagnets.
\end{abstract}
\maketitle

\section{Introduction}
\label{chap:skyrmion:sec:intro}

For systems with broken symmetries, low-energy effective theories 
provide a simple and powerful tool for describing the relevant
physics.  The focus is on
the long-wavelength Goldstone modes which emerge from the microscopic 
degrees of freedom.  This often reveals the connections between
seemingly unrelated systems that share the same pattern of symmetry
breaking.  The most well-known example is that of the complex
scalar field used in the low-energy theories of bosonic 
superfluids \cite{gross-61,pitaevskii-61},
fermionic superconductors \cite{ginzburg-50}, 
and XY spin systems \cite{kosterlitz-73}.

Effective theories also bring to light subtle topological
effects that may become important at low energies.  Goldstone modes
often carry a non-trivial topology which can give rise
to the appearance of topological defects.
For example, in two spatial dimensions
Kosterlitz and Thouless pointed out the crucial role of vortices 
for the complex scalar field \cite{kosterlitz-73,minnhagen-87}.
Vortices are point-like topological defects around which
the phase of the complex scalar field winds by an integer multiple of
$2\pi$.

Ultracold atomic systems provide an ideal testing ground for low-energy
effective theories.  The microscopic degrees of freedom are well-isolated
from the environment experimentally and well-understood theoretically.
The challenge is in describing how these microscopic degrees of freedom
organize at low energies in the presence of non-trivial interactions.
When only one internal hyperfine level is important, the phenomenon
of scalar Bose-Einstein condensation is given by 
the low-energy theory of the complex scalar field 
\cite{stringari-book,smith-book,davidson-05}.
A series of ground breaking experiments have observed
vortex lattices in rotating condensates
\cite{cornell-99,ketterle-01} as well as 
evidence for the role of vortices in the equilibrium 
Kosterlitz-Thouless transition for two-dimensional 
condensates \cite{hadzibabic-06}.

For ultracold atoms with a complex
internal level structure, there are various patterns of symmetry breaking.  This
lead to rich possibilities and challenges for low-energy effective 
descriptions.
Spinor condensate ferromagnets are one such system which has seen a number
of important experimental advancements (see Ref. \cite{stamper-kurn-00}).
This includes the development of optical
dipole traps used for preparation \cite{mit-98a} and phase-contrast imaging used for
detection \cite{berkeley-05} in $S=1$ $^{87}$Rb.  
In addition to the phase degree of freedom
familiar from single-component condensates, 
the magnetization naturally arises as a description of
the low-energy spin degrees of freedom.  A vector quantity sensitive to both
the population and coherences between the three hyperfine levels,
the magnetization can be directly imaged in experiments \cite{berkeley-05}.

One of the most striking observations in spinor condensate ferromagnets
is the spontaneous formation of crystalline magnetic 
order \cite{dsk-08,dsk-09}.
From an initial quasi-two-dimensional condensate prepared with a uniform 
magnetization, a crystalline lattice of spin domains emerges spontaneously
at sufficiently long times.  
The presence of a condensate with magnetization spontaneously breaks 
global gauge invariance 
and spin rotational invariance.
Additionally, crystalline order for the magnetization breaks 
real space translational and rotational symmetry.
Previous works have pointed out the crucial
role of dipolar interactions in driving dynamical instabilities within
the uniform condensate towards states with crystalline order 
\cite{cherng-09,ueda-09,sau-09}.
Numerical analysis of the full multi-component
Gross-Pitaevskii equations suggest dipolar interactions can give rise
to states with crystalline order \cite{ueda-09,ho-09}.

In this and a companion paper \cite{cherng-10b}, we take a complementary approach 
and focus on the low-energy effective theory of two-dimensional 
spinor condensate ferromagnets.
This effective theory describes the interaction between the superfluid velocity and the magnetization degrees of freedom.
Previous work has derived the equations of motion for this effective 
theory \cite{lamacraft-07b,barnett-09}.
We extend this result to demonstrate how the Lagrangian for the effective theory 
can be written as a non-linear sigma model in terms of the magnetization alone.
The effect of the superfluid velocity is to induce a long-ranged interaction term 
between skyrmions, topological defects familiar from the theory of ordinary 
ferromagnets \cite{rajamaran-82}.
In contrast to the point-like topological defects of vortices,
skyrmions describe extended magnetization
textures which carry a quantized topological charge.

In the companion paper \cite{cherng-10b}, 
we study how symmetry groups containing combined
real space translational, real space rotational, and spin space rotational
symmetry operations can be used to classify possible crystalline magnetic orders.
Within each symmetry class, we find minimal energy configurations
describing non-trivial crystalline configurations.
 
The main purpose of this paper is to give a simple physical physical picture
for the origin of crystalline order
in terms of neutral configurations of skyrmions and anti-skyrmions.
Long-ranged skyrmion interactions
force magnetization configurations to have
net neutral collections of topological defects.
We show
this explicitly by finding exact analytical solutions for the non-linear 
equations of motion describing both localized collections of skyrmions
and anti-skyrmions as well as extended skyrmion and anti-skyrmion 
stripes.

Skyrmions have non-trivial spin configurations that spontaneously break
translational and rotational invariance in real space.
Proposed originally in high energy physics as a model for mesons and baryons
\cite{skyrme-62,klebanov-85}, skyrmions have found applications in a number
of diverse fields including quantum hall ferrogmagnets \cite{sondhi-93,brey-95},
and magnetically ordered crystals 
\cite{yablonskii-89,muhlbauer-09,neubauer-09,rosch-09}.

A neutral collection of 
such topological objects is able to take advantage of the dipolar
interaction energy without a large penalty in the skyrmion interaction energy.
Since a neutral skyrmion configuration has the same topological number as the
uniform magnet, its stability is not ensured by topology alone.  However,
the scale invariance of the skyrmion interaction energy rules out the most 
straightforward instability of bringing skyrmions and anti-skyrmions closer 
together.  Essentially, the skyrmion interaction energy forces the charge 
densities of a skyrmion and anti-skyrmion pair to shrink as the distance between
them shrinks so that the overall energy remains the same.

The analytical solutions we find without dipolar interactions
closely resemble the minimal energy configurations found numerically in the presence
of dipolar interactions.
The role of dipolar interactions can then be
seen as stabilizing these non-trivial solutions of the effective theory.
We point out that magnetic dipolar interactions are small. 
Thus it is a good starting point to find states which are static 
nontrivial solutions of the system without dipolar interactions.

The effective theory of spinor condensate ferromagnets
is essentially identical to that of the quantum Hall 
ferromagnets \cite{sondhi-93,kane-90}.
Although the microscopic degrees of freedom are fermionic electrons, the
magnetization order parameter is bosonic.  The Coloumb interaction between
electrons then gives a contribution to the skyrmion interaction for the 
effective theory.  The resulting skyrmion interaction is qualitatively the same
as the one for spinor condensate ferromagnets.  This suggests the study 
of spinor condensate ferromagnets may have interesting connections
to quantum hall ferromagnets and vice versa.

The plan of this paper is as follows.  In Sec. \ref{chap:skyrmion:sec:ferromagnets}
we review the theory of ordinary ferromagnets and how skyrmion solutions
arise from the non-linear Landau-Lifshitz equations of motion.
We then proceed to review how these skyrmion solutions are used in
the low-energy effective theory of quantum Hall ferromagnets in
Sec. \ref{chap:skyrmion:sec:qh}.  Next we derive the low-energy effective theory
for spinor condensate ferromagnets in Sec. \ref{chap:skyrmion:sec:spinor}.  In particular,
we demonstrate how a long-ranged skyrmion interaction term (which also
appears for quantum hall ferromagnets) arises
from coupling of the magnetization to the superfluid velocity.

In Sec. \ref{chap:skyrmion:sec:exact}, we discuss how to interpret the mathematical structure
of skyrmion solutions for ordinary ferromagnets in terms of a separation
of variables.  This approach allows us to use find new exact solutions for
the oridinary ferromagnet.  More importantly, it also allows us 
to generalize the skyrmion solutions of the ordinary ferromagnet to
find analytical solutions for the spinor condensate ferromagnet.
These latter solutions describe both neutral collections
of localized skyrmions and anti-skyrmions as well as extended stripe
configurations.  Finally, we discuss how the analytical
solutions we find offer insight into quantum hall ferromagnets and 
spinor condensate ferromagnets with dipolar interactions in 
Sec. \ref{chap:skyrmion:sec:discussion}

\section{Skyrmions in ferromagnets}
\label{chap:skyrmion:sec:ferromagnets}

We begin by reviewing the theory of ordinary two-dimensional ferromagnets
described by the following Lagrangian, Hamiltonian, and 
Landau-Lifshitz equations of motion \cite{fradkin-91}
\begin{align}
\nonumber
\mathcal{L}&=
-S\int dt d^2x \mathcal{A}(\hat{n})\cdot\partial_t\hat{n}-\int dt \mathcal{H}\\
\nonumber
\mathcal{H}&=\frac{S}{4}\int d^2x\nabla(\hat{n})^2\\
\partial_t\hat{n}&=
\frac{1}{2}\hat{n}\times\nabla^2\hat{n}
\label{chap:skyrmion:eq:ferromagnet_lagrangian}
\end{align}
where $\hat{n}$ is a three component real unit vector
and $\mathcal{A}(\hat{n})$ is the unit monopole vector potential.
The order parameter $\hat{n}$ describes the magnetization
and is a unit vector living on the sphere.
Calculating the variation of the Lagrangian to derive the
Landau-Lifshitz equations \cite{stone-89,stone-96} can be done by using
$\delta \hat{n}=\delta w\times \hat{n}$.
This is consistent with the constraint $|\hat{n}|=1$
since $\delta\hat{n}\cdot \hat{n}=0$ by construction.
The variation of the term 
$\int dt \mathcal{A}(\hat{n})\cdot\partial_t\hat{n}$ 
is clearest when using its geometric interpretation as
the area on the sphere swept by $\hat{n}$.
While we include time dependent terms for completeness,
in this paper we will only consider static solutions

In addition to the trivial uniform solution, 
there are non-trivial soliton solutions to the Landau-Lifshitz equations
called skyrmions.  By parameterizing
\begin{align}
\hat{n}=\begin{bmatrix}
\sin(\alpha)\cos(\beta)&
\sin(\alpha)\sin(\beta)&
\cos(\alpha)
\end{bmatrix}^T
\label{chap:skyrmion:eq:skyrmion}
\end{align}
we see $\alpha$ controls $\hat{n}_z$, the $\hat{z}$ component of the
magnetization while $\beta$ controls the canonically
conjugate variable giving the orientation of $\hat{n}_x$,
$\hat{n}_y$, the $\hat{x}$, $\hat{y}$ components of
the magnetization.

The minimal energy solutions within each topological sector are called skyrmions
and can be written in the form \cite{rajamaran-82}
\begin{align}
\label{chap:skyrmion:eq:skyrmion_solution}
\tan(\alpha/2)e^{i\beta}=\exp[f(x+iy)]=\exp[u(x,y)+iv(x,y)]
\end{align}
where $f(z)$ is a holomorphic function of $z$ with real part
$u(x,y)$ and imaginary part $v(x,y)$.
The function $f(z)\sim n\log(z-z_0)$ can have logarithmic singularities
\footnote{The convention in literature on 
skyrmions in ordinary ferromagnets 
is to take $\tan(\alpha/2)e^{i\beta}=g(x+iy)$ with $g(z)$ allowed to
have isolated zeros or poles.  Taking $f(z)=\log g(z)$ and allowing
$f(z)$ to have logarithmic singularities is equivalent and can be more easily
generalized to spinor condensate ferromagnts.}.
Since $\beta$ has to be $2\pi$ periodic, the residues
of these singularities must be integers.
This implies $\exp[f(x+iy)]$ can only have zeros or poles.
The resulting spin configuration has $\hat{n}_z=+1$ 
($\hat{n}_z=-1$) at zeros (poles) of $\exp[f(z)]$ while 
$\hat{n}_x$, $\hat{n}_y$
wind anti-clockwise (clockwise) along a path circling the zeros 
(poles) in a anti-clockwise direction.

These solutions describe topological defects of ordinary 
ferromagnets.
For fixed boundary conditions, smooth, finite energy configurations are 
separated into distinct classes characterized by a quantized topological
invariant $\int d^2x q(x)=4\pi N$ where $N$ is the number of times $\hat{n}$ 
covers the sphere.  Here
\begin{align}
q=\epsilon_{\mu\nu} \hat{n}\cdot \nabla_\mu \hat{n}\times \nabla_\nu\hat{n}
\label{chap:skyrmion:eq:skyrmion_density}
\end{align}
is the skyrmion density which is positive for $f(z)$ holomorphic.
From here on, lower Greek indices (upper Roman indices) refer to
real space (order parameter) components.

\begin{figure}
\begin{center}
\begin{tabular}{cc}
\includegraphics[width=1.5in]{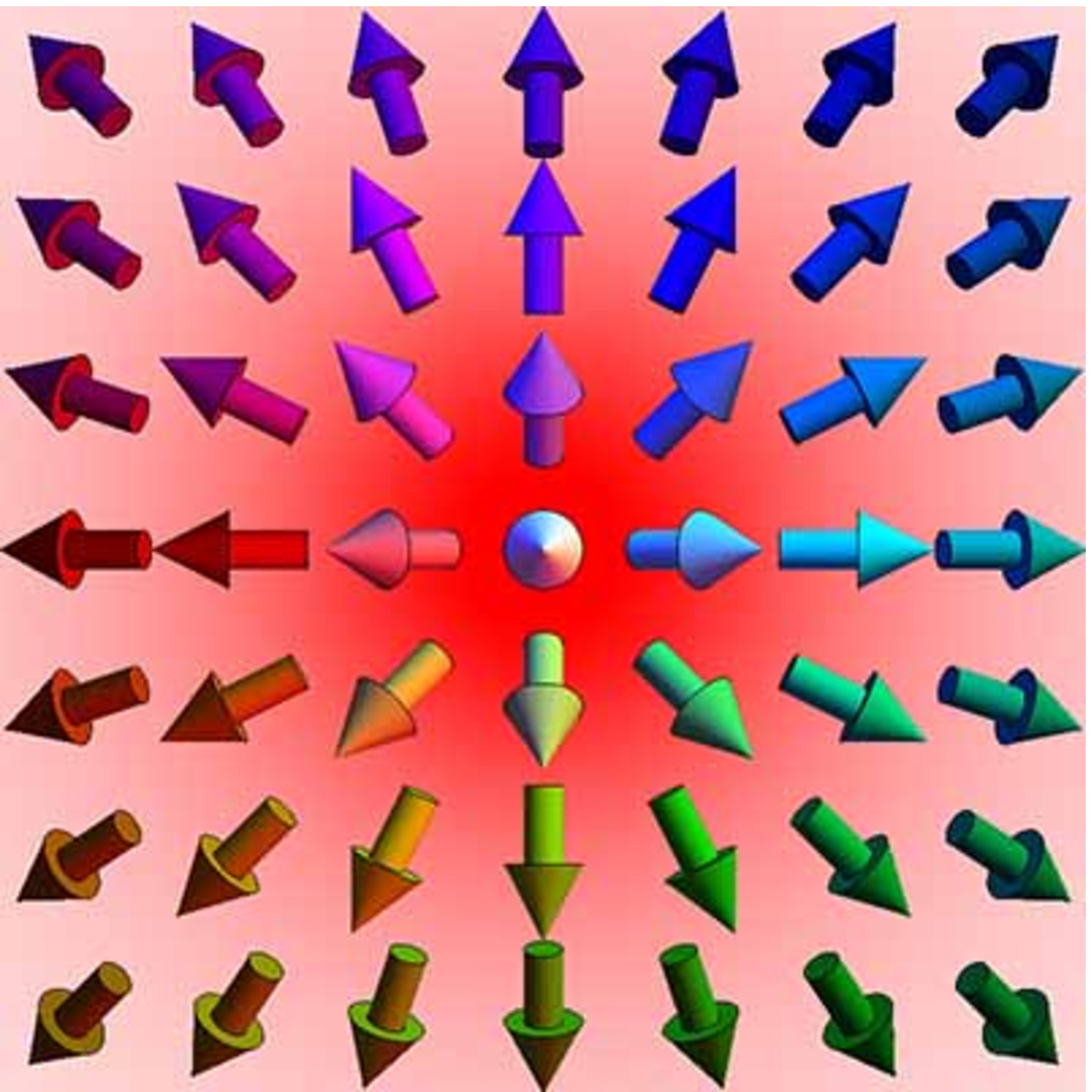}&
\includegraphics[width=1.5in]{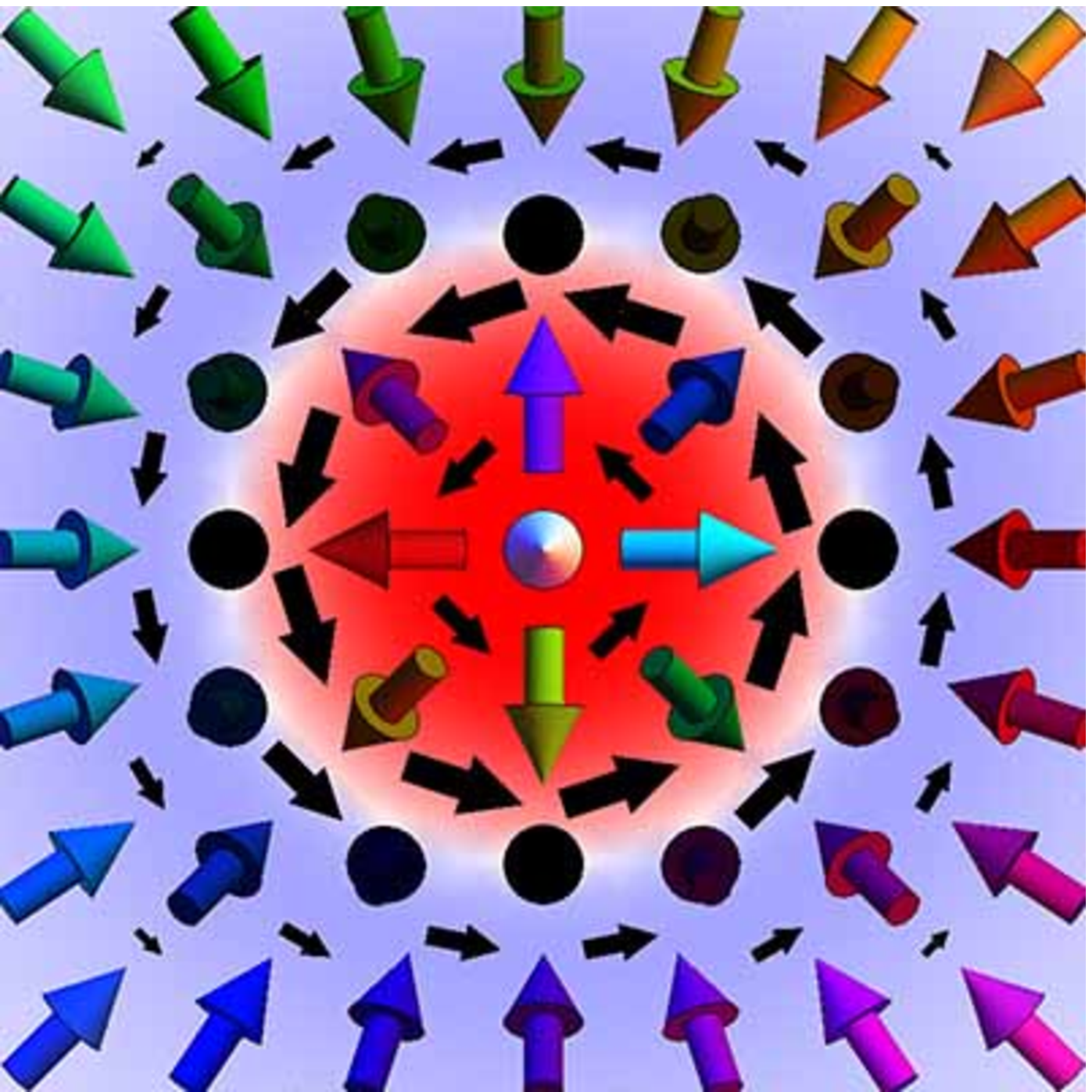}
\end{tabular}
\end{center}
\caption[Localized skyrmions and anti-skyrmions]{
A single localized skyrmion carrying $4\pi$ net skyrmion charge
in the ordinary ferromagnet (left).
Neutral configuration consisting of a localized skyrmion carrying
$+4\pi$ skyrmion charge in a negative background carrying $-4\pi$
skyrmion charge in the spinor condensate ferromagnet (right).
Notice the $+\hat{z}$ ($-\hat{z})$ meron
carrying $+2\pi$ ($+2\pi$) net skyrmion charge
at the origin (at infinity) for the ordinary ferromagnet
are mapped to a skyrmion (anti-skyrmion)
carrying $+4\pi$ (-$4\pi$) net skyrmion charge
in the spinor condensate ferromagnet.
Red (blue) background indicates positive (negative) skyrmion 
density $q$, black 2D arrows the superfluid velocity $\mathbf{v}$ for
the spinor condensate ferromagnet, 
and shaded 3D arrows the magnetization $\hat{n}$.
}
\label{chap:skyrmion:fig:skyrmion_single}
\end{figure}

Consider the single skyrmion solution shown in the left of
Fig. \ref{chap:skyrmion:fig:skyrmion_single}.
It corresponds to $f(z)=\log(z)$, carries net skyrmion charge $4\pi$,
and can be decomposed into a $+\hat{z}$ meron at the origin and a
$-\hat{z}$ meron at infinity.  Essentially, a meron can be thought of
as half of a skyrmion and characterized by two signed quantities: the direction
of the magnetization at the core and the orientation of the winding away from
the core.  The sign of the skyrmion density is the product of the sign of these 
two quantities.  For example, the $+\hat{z}$ meron at the origin to the left
of Fig. \ref{chap:skyrmion:fig:skyrmion_single} carries 
net skymrion charge $2\pi$.

For spinor condensate ferromagnets, scale invariance of the skyrmion interaction 
term guarantees stability against trivial rescaling.  However, this may
change if we introudce a short distance cutoff or quantum fluctuations.

Notice the Lagrangian 
in Eq. \ref{chap:skyrmion:eq:ferromagnet_lagrangian} has translational, rotational, and scale
invariance in real space as well as 
rotational invariance in spin space.
The single skyrmion solution spontaneously breaks all of these symmetries.
However, the action is invariant for $f(z)=\log[f_0(z-z_0)]$ with 
complex constants $z_0$ and $f_0$.
Taking solutions with $z_0\ne0$ corresponds to spatially translating
the solution with $z_0=0$,
solutions with $\text{Re}[f_0]\neq 1$ 
correponds to spatially rescaling the the solution with $f_0=1$,
and solutions with $\text{Im}[f_0]\neq 0$ corresponds to rotation of the 
$f_0=1$ solution. 
In addition, the action is the same
for $O\hat{n}$ where $O$ is a rotation matrix describing solutions related
by spin space transformations.

\begin{figure}
\begin{center}
\begin{tabular}{cc}
\includegraphics[width=1.5in]{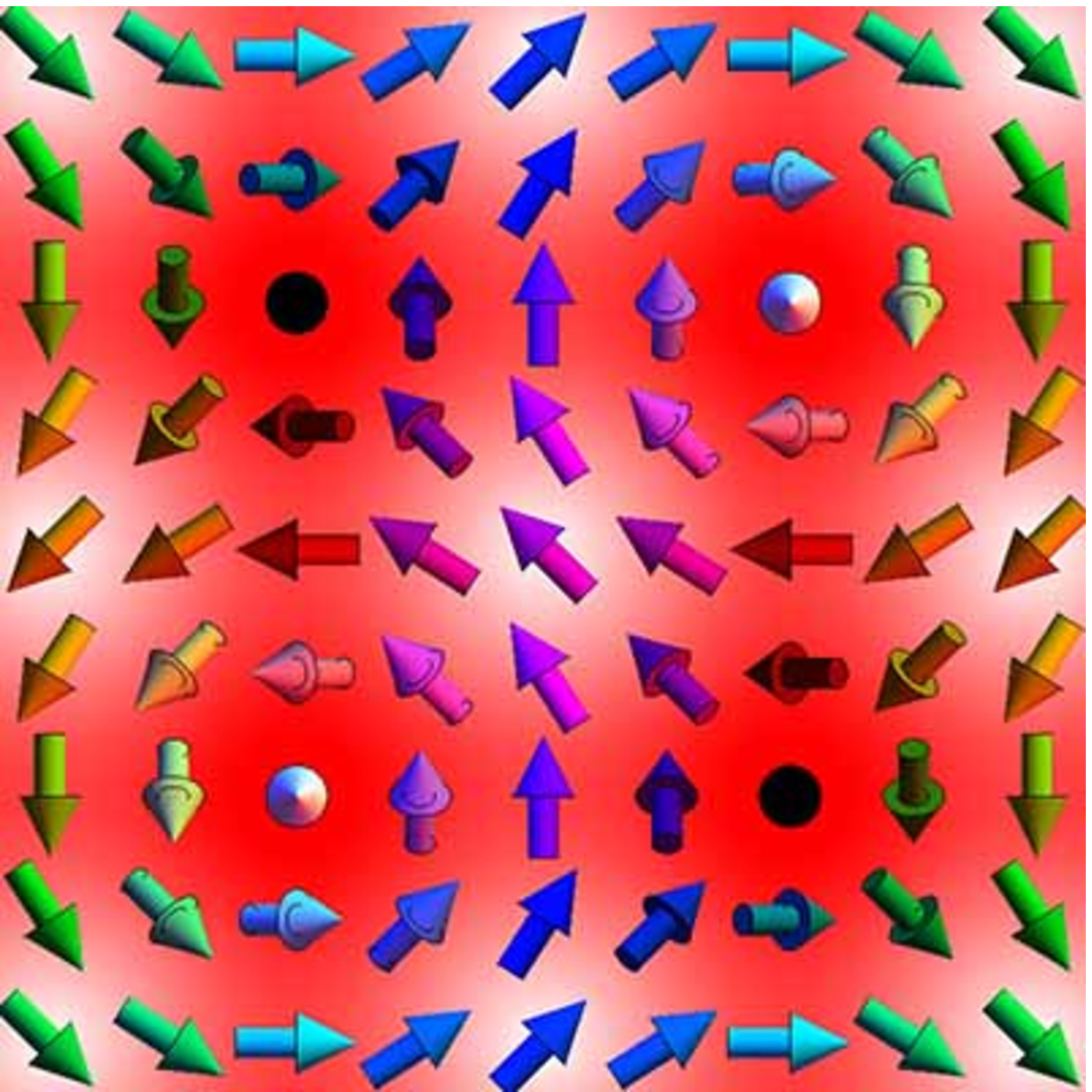}
\includegraphics[width=1.5in]{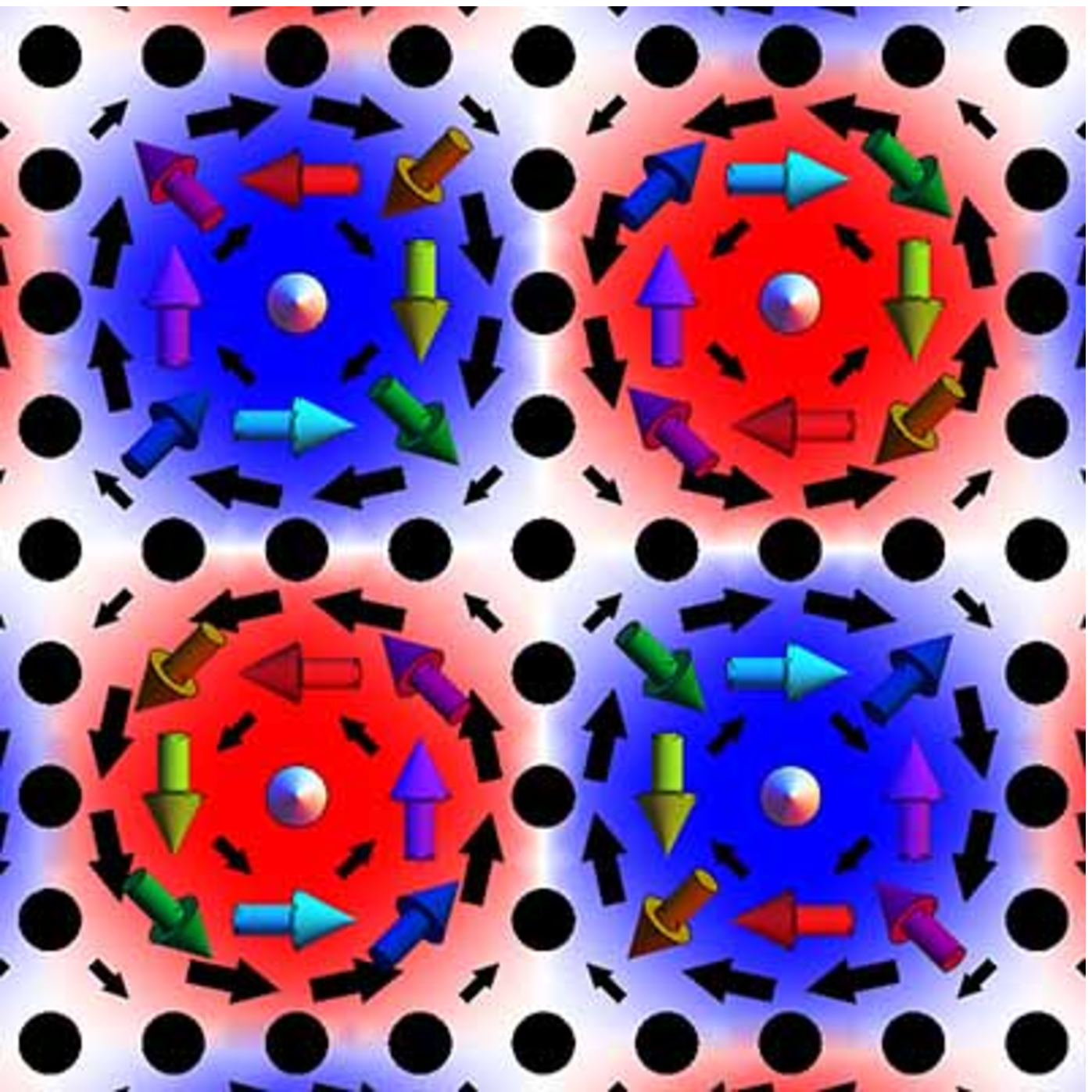}
\end{tabular}
\end{center}
\caption[Lattice of skyrmions and anti-skyrmions]{
Unit cell for a lattice of localized skyrmions carrying $8\pi$ 
net skyrmion charge per unit cell
in the ordinary ferromagnet (left).
Unit cell for a neutral configuration carrying
zero net skyrmion charge per unit cell
in the spinor condensate ferromagnet (right).
Notice $+\hat{z}$ ($-\hat{z})$ merons
are mapped to skyrmions (anti-skyrmions).
Red (blue) background indicates positive (negative) skyrmion 
density $q$, black 2D arrows the superfluid velocity $\mathbf{v}$, 
and shaded 3D arrows the magnetization $\hat{n}$.
}
\label{chap:skyrmion:fig:skyrmion_lattice}
\end{figure}

Collections of multiple skyrmions are described by $f(z)$ having multiple
singularities.  A square lattice of skyrmions with net skyrmion charge
$8\pi$ per unit cell is shown in the left of Fig. \ref{chap:skyrmion:fig:skyrmion_lattice}.
It can also be decomposed into a collection of two $+\hat{z}$ merons
and two $-\hat{z}$ merons arranged antiferromagnetically.
In addition, there are solutions
$\tan(\alpha/2)e^{i\beta}=\exp[f(x-iy)]$ with $f(\bar{z})$ antiholomorphic
characterized by  $\hat{x}$, $\hat{y}$ components winding in the opposite
direction compared to holomorphic solutions and $q$ negative.

\section{Quantum Hall Ferromagnets}
\label{chap:skyrmion:sec:qh}

After describing the structure of skyrmion solutions for ordinary ferromagnets,
we now briefly review how they are used in the study of quantum Hall 
ferromagnets \cite{sondhi-93,kane-90}.
For the low-energy effective theory of quantum Hall ferromagnets,
spin $1/2$ electrons in a magnetic field are described
by a two component Chern-Simons composite boson theory.
The bosons couple to the physical gauge field due to the magnetic field
and a fictitious Chern-Simons gauge field which attaches
appropriate flux quanta to change from bosonic to fermionic statistics.
At appropriate filling fractions, the physical and Chern-Simons fluxes
cancel and the resulting quantum Hall plateau is described as a 
condensate of composite bosons.

The Lagrangian and Hamiltonian for the resulting effective theory is given by
\begin{align}
\nonumber
\mathcal{L}=&
-\frac{1}{2}\int dt d^2x \mathcal{A}(\hat{n})\cdot\partial_t\hat{n}
-\int dt \mathcal{H}\\
\nonumber
\mathcal{H}=&\int d^2x\left[
\frac{1}{8}(\nabla\hat{n})^2+\frac{1}{2}\vec{B}\cdot \hat{n}\right]\\
&+\frac{1}{8}\int d^2x d^2y[q(x)-\bar{q}]G(x-y)[q(y)-\bar{q}]
\label{chap:skyrmion:eq:lagrangian_qh}
\end{align}
where $\vec{B}$ is the magnetic field giving rise to a linear Zeeman 
shift, $q(x)$ is the skyrmion density
and $\bar{q}=\int d^2x q(x)$ is the net skyrmion charge.

The coupling of the bosons to the Chern-Simons field gives rise to the
skyrmion interaction term.  At long distances, the Chern-Simons field
ties the skyrmion density to the deviation of the physical electron
density from its background value.
In fact skyrmions are lowest energy quasiparticles at a filling factor of
one \cite{sondhi-93,fertig-94}.
Thus, the skyrmion density 
inherits the Coloumb interaction for the electron density.
Away from quantum Hall plateaus, the
background value $\bar{q}$ is non-zero and the singular 
$|x-y|^{-1}$ behavior in 
$G(x-y)$ forces finite energy configurations to have
non-zero net skyrmion charge.

Thus the focus is on static configurations carrying net 
skyrmion charge.
The skyrmion solutions of the ordinary ferromagnet have non-zero $\bar{q}$
and provide a good qualitative description of quantum Hall ferromagnets
away from quantum Hall plateaus.
For example, a sqaure lattice of skyrmions as shown in Fig. 
\ref{chap:skyrmion:fig:skyrmion_lattice} carries a net charge of $8\pi$ per unit cell
and can be used as a starting point for studying configurations carrying finite
$\bar{q}$.
Although such solutions take into account the spin stiffness and the
long-ranged divergence of the skyrmion interaction, quantitative
results require additional analysis arising from the 
detailed form of skyrmion interaction and the linear Zeeman shift.
This usually involves numerical 
minimization \cite{sondhi-97a,sondhi-99,fertig-94,girvin-98} of the Hamiltonian in
Eq. \ref{chap:skyrmion:eq:lagrangian_qh}.

\section{Spinor Condensate Ferromagnets}
\label{chap:skyrmion:sec:spinor}

Having reviewed known results on ordinary and quantum Hall ferromagnets,
we now consider spin $S$ spinor condensate ferromagnets
described by the microscopic condensate wavefunction $\mathbf{\Psi}$,
a $2S+1$ complex vector \cite{machida-98,ho-98}.
The microscopic Gross-Pitaevskii Lagrangian is given by
\begin{align}
\nonumber
\mathcal{L}&=\int dt i\mathbf{\Psi}^\dagger\partial_t\mathbf{\Psi}
-\int dt \mathcal{H}-\int dt \mathcal{H}_S\\
\mathcal{H}&=\int d^2x\left[
\frac{1}{2m}|\nabla\mathbf{\Psi}|^2
+g_0(\mathbf{\Psi}^\dagger\mathbf{\Psi})^2
+g_s(\mathbf{\Psi}^\dagger\vec{F}\mathbf{\Psi})^2
\right]
\label{chap:skyrmion:eq:gp}
\end{align}
where $\vec{F}$ are spin matrices and
$g_0>0$ gives the spin-independent contact interaction strength
while $g_s<0$ gives the spin-dependent contact interaction strength
which favors finite magnetization.
Here, $\mathcal{H}_S$ denotes additional spin dependent interactions such 
as the quadratic Zeeman shift and dipolar interactions.

As was the case for quantum Hall ferromagnets, 
we expect a simpler description to emerge
at low energies.
The resulting low-energy effective theory should only involve
the condensate phase $\phi$ and local magnetization $\hat{n}$ which
describe the order parameters of the system.
Previous work has shown this can be done at the level of the equations
of motion \cite{lamacraft-07b}.
In this section, we extend this result to derive the Lagrangian 
and Hamiltonian for the effective theory solely in terms of the magnetization.
However, the skyrmion density acts as a source of voriticity for
the superfluid velocity.   
Thus, the effect of the superfluid phase is to induce 
a logarithmic vortex-vortex interaction between skyrmions.
The resulting non-linear sigma model is essentially identical to
that of the quantum Hall ferromagnet
but with skyrmion interaction 
$G(x-y)\sim\log(x-y)$ having logarithmic behavior instead of 
$|x-y|^{-1}$ behavior.  

We begin by considering energies below the scale of spin-independent $g_0$
and ferromagnetic spin-dependent $g_s$
contact interactions.  The condensate has fixed density
$\mathbf{\Psi}^\dagger\mathbf{\Psi}=\rho$ and fully polarized magnetization
$\mathbf{\Psi}^\dagger \vec{F} \mathbf{\Psi}=S\rho \hat{n}$ where
$\vec{F}$ are spin matrices.
The states that satisfy these constraints are parameterized solely
in terms of the low-energy degrees of freedom
\begin{align}
\mathbf{\Psi}&=\sqrt{\rho}e^{i\phi}\psi_{\hat{n}},&
\hat{n}\cdot\vec{F}\psi_{\hat{n}}&=S\rho\psi_{\hat{n}}
\end{align}
where $\phi$ descrribes the phase of the condensate
and $\psi_{\hat{n}}$ is a fully polarized unit spinor
with $\hat{n}$ describing the orientation of the magnetization.
Although $\phi$ is not directly observable, the 
superfluid velocity is a physical quantity
\begin{align}
\mathbf{v}_\mu=\nabla_\mu\phi-i\psi_{\hat{n}}^\dagger\nabla_\mu\psi_{\hat{n}}
\label{chap:skyrmion:eq:sf}
\end{align}
which has contributions from both $\phi$ and $\psi_{\hat{n}}$.

In this paper, we are primarily interested in the
competition between the spin stiffness and superfluid
kinetic energy.  In the companion paper \cite{cherng-10b}, we address
the effect of magnetic dipolar interactions.  From here on,
we consider the case $\mathcal{H}_S=0$.
From the Gross-Pitaevskii Lagrangian in Eq. \ref{chap:skyrmion:eq:gp},
the Berry's phase term becomes
$i\mathbf{\Psi}^\dagger\partial_t\mathbf{\Psi}=
-\rho\partial_t\phi
-S\rho\mathcal{A}(\hat{n})\cdot\partial_t\hat{n}$
while the kinetic energy term is given by
$|\nabla\mathbf{\Psi}|^2=S\rho/2(\nabla\hat{n})^2+\rho\mathbf{v}^2$
and the interaction terms give constants.
This gives the Lagrangian and Hamiltonian as
\begin{align}
\nonumber
\mathcal{L}&=
-S\int dt d^2x \mathcal{A}(\hat{n})\cdot\partial_t\hat{n}
-\int dt \mathcal{H}\\
\mathcal{H}&=\int d^2x 
\left[\frac{S}{4}(\nabla\hat{n})^2
+\frac{1}{2}\mathbf{v}^2
\right]
\label{chap:skyrmion:eq:spinor_lagrangian}
\end{align}
where we take $\rho=m=1$ for simplicity.
Notice for fixed $\rho$, the $\partial_t\phi$ term is a total 
derivative which we exclude.
Compared to the Lagrangian describing ordinary
ferromagnets in Eq. \ref{chap:skyrmion:eq:ferromagnet_lagrangian},
there is an additional superfluid kinetic energy term
$\mathbf{v}_\mu\mathbf{v}_\mu$.

The global phase $\phi$ enters the Lagrangian quadratically
and only through $\mathbf{v}$.  The equation of motion for
$\phi$ gives $\nabla_\mu \mathbf{v}_\mu=0$
implying the superfluid velocity is divergenceless.
This follows from $\mathbf{v}$ describing transport of
the density $\rho$, a conserved quantity which is locally fixed
at low energies due to the spin-independent contact interaction.
This implies that $\mathbf{v}_\mu\mathbf{v}_\mu$
only depends on the divergenceless part of $\mathbf{v}$.
In momentum space, this is
$\mathbf{v}_\mu(+k)[\delta_{\mu\nu}-k_\mu k_\nu/k^2]\mathbf{v}_\nu(-k)$
which can also be written as $F_{\mu\nu}(+k)F_{\mu\nu}(-k)/2k^2$.

Here we have introduce the analog of the field strength tensor
$F_{\mu\nu}=\nabla_\mu \mathbf{v}_\nu-\nabla_\nu \mathbf{v}_\mu$, a local quantity
that depends only on the divergenceless part of $\mathbf{v}$.
In two dimensions, there is only 
one non-zero component to $F_{\mu\nu}$. 
From Eq. \ref{chap:skyrmion:eq:sf}, this is given by the skyrmion 
density $F_{xy}=-F_{yx}=Sq$.
Here we assume that the condensate phase $\phi$
does not contribute to $F_{xy}$ through vortex-like
singularities.  This is valid because the vortex core energy
is large.

The above results give $\mathbf{v}$ 
solely in terms of $\hat{n}$ as
\begin{align}
\nabla_\mu \mathbf{v}_\mu&=0,&
\epsilon_{\mu\nu} \nabla_\mu \mathbf{v}_\nu&=Sq
\label{chap:skyrmion:eq:superfluid_eom}
\end{align}
with $q$ the skyrmion density.
Gradients in the order parameter $\hat{n}$ arise in part from 
phase gradients in the condensate wavefunction $\mathbf{\Psi}$.
Topologically non-trivial magnetization configurations
can thus give rise to vorticity described 
by a non-zero curl $\epsilon_{\mu\nu}\nabla_\mu\mathbf{v}_\nu\ne0$.

By introducing the two-dimensional logarithmic Green's function 
$-\nabla^2 G(x)=\delta(x)$ we can write the superfluid kinetic
energy $F_{\mu\nu}(+k)F_{\mu\nu}(-k)/2k^2$ in real space and obtain
\begin{align}
\nonumber
\mathcal{L}&=
-S\int dt d^2x \mathcal{A}(\hat{n})\cdot\partial_t\hat{n}
-\int dt \mathcal{H}\\
\mathcal{H}&=\frac{S}{4}\int d^2x 
(\nabla\hat{n})^2
+\frac{S^2}{2}\int d^2x d^2y q(x)G(x-y)q(y)
\label{chap:skyrmion:eq:final_lagrangian}
\end{align}
with the corresponding equations of motion given by
\begin{align}
\left(\partial_t+\mathbf{v}_\mu\nabla_\mu\right)\hat{n}&=
\frac{1}{2}\hat{n}\times\nabla^2\hat{n}
\label{chap:skyrmion:eq:eom}
\end{align}
along with Eq. \ref{chap:skyrmion:eq:superfluid_eom} for the superfluid velocity solved by
\begin{align}
\mathbf{v}_\mu&=S\epsilon_{\mu\nu}\nabla_\nu\Phi,&
-\nabla^2\Phi&=q
\label{chap:skyrmion:eq:superfluid}
\end{align}
where $\Phi(x)=\int d^2x G(x-y)q(y)$ has the interpretation
of the two-dimensional Coloumb potential associated with $q$.

Compared to the Landau-Lifshitz equations describing ordinary
ferromagnets in Eq. \ref{chap:skyrmion:eq:ferromagnet_lagrangian},
the replacement $\partial_t\rightarrow \partial_t+\mathbf{v}_\mu\nabla_\mu$
describes the advection of the magnetization by the superfluid velocity
\cite{lamacraft-07b}.
This advective term arise from variation of the superfluid kinetic energy
term in Eq. \ref{chap:skyrmion:eq:spinor_lagrangian} or equivalently from 
the skyrmion interaction term in Eq. \ref{chap:skyrmion:eq:final_lagrangian}.
Recall that we include the time dependence for completeness and
focus only on static solutions.

The skyrmion density $q$ gives the vorticity for 
the superfluid velocity $\mathbf{v}$.  Thus, the second term in the Hamiltonian
above gives the pairwise logarithmic interaction energy between vortices.
In the thermodynamic limit, the logarithmic divergence of $G(x-y)$
at large distances forces finite energy configurations to
have zero net skyrmion density $\int d^2x q(x)=0$.

Notice Eq. \ref{chap:skyrmion:eq:final_lagrangian} for spinor condensate
and Eq. \ref{chap:skyrmion:eq:lagrangian_qh} for quantum hall ferromagnets
have the same form.  Although $G(x-y)$ behaves as $\log|x-y|$ for
the former and $|x-y|^{-1}$ for the latter, both give singular
contributions
at long wavelengths.  However, the important absence of 
a finite backgroud value $\bar{q}$ in the skyrmion interaction 
implies configurations for spinor condensate ferromagnets must have
zero net skyrmion charge.

\section{Exact solutions with neutral skyrmion charge}
\label{chap:skyrmion:sec:exact}

In quantum hall systems density deviations from the incompressible state 
cause skyrmions. 
Density fluctuations with zero net average (such as impurities) cause spin 
textures with zero net skyrmion number. States with non-zero skyrmion
number occur away from quantum hall plateaus.
Thus the analytical skyrmion solutions carrying 
\textit{net charge} for the ordinary ferromagnet offered insight
into more complicated case of 
quantum Hall ferromagnets away from the quantum Hall plateau.
For spinor condensate ferromangets, 
we will show how \textit{net neutral} solutions with skyrmions
and anti-skyrmions without dipolar interactions offer insight into
the more complicated case with dipolar interactions.

In this section, we find exact analytical solutions for spinor condensate
ferromagnets with logarithmic skyrmion interactions in the absence 
of
dipolar interactions.
We study the effect of including dipolar interactions 
numerically after a symmetry analysis in the companion paper \cite{cherng-10b}.
The exact solutions we find here greatly resemble the numerical solutions in the
companion paper.  As we discuss in Sec. \ref{chap:skyrmion:sec:discussion}, 
the interpretation of the exact solutions
in terms of neutral collections of skyrmions and anti-skyrmions offers
physical insight into the more complicated numerical solutions 
of the companion paper.

To find exact solutions with zero net skyrmion charge, 
it is vital to include to effect of the skyrmion interaction term.
Recall it is the long wavelength divergence 
of this term that forces configurations
to have zero net skyrmion charge.
Although this cannot be done exactly for $|x-y|^{-1}$ interactions as
in quantum hall ferromagnets, it is possible for logarithmic 
interactions as in spinor condensate ferromagnets.   Physically, 
this is because the logarithmic interaction arises solely from the 
superfluid kinetic energy which is scale invariant just like the spin stiffness term.

We begin with the parameterization of $\hat{n}$ in 
Eqs. \ref{chap:skyrmion:eq:skyrmion}, \ref{chap:skyrmion:eq:skyrmion_solution},
used in the skyrmion solutions of the ordinary ferromagnet.  Notice
$\alpha$, $\beta$ provide a set of orthogonal coordinates for the sphere
describing the order parameter space of $\hat{n}$.
For $f(x+iy)=u(x,y)+iv(x,y)$ holomorphic,
$u(x,y)$ and $v(x,y)$ provide a set of orthogonal coordinates for the plane
describing real space.
So for ordinary ferromagnets, skyrmion solutions are given by 
$\alpha=2\tan^{-1}(e^u)$ and $\beta=v$ 
(see also Eq. \ref{chap:skyrmion:eq:skyrmion_solution}), which can be understood as
a separation of variables.

Notice $\alpha(u)$ is a function of $u$ only while $\beta(v)$ is a function
of $v$ only.
Each orthogonal coordinate of the order parameter space 
$\alpha$, $\beta$ is a function of only one orthogonal coordinate of 
real space $u$, $v$, respectively.
The reason why using $u$ and $v$ as coordinates is tractable 
is because they satisfy the Cauchy-Riemann equations
$\partial_x u=+\partial_y v$, $\partial_y u=-\partial_x v$.
In particular, this implies $\nabla u\cdot \nabla v=0$ meaning
countour lines of constant
$u$ are perpendicular to countour lines of constant $v$ as required
for orthogonal coordinates.  In addition, both $\nabla^2 u=0$ and $\nabla^2 v=0$
satisfy Laplace's equation.  The above two identities simplify
expressions involving $\nabla^2$ which arise in the 
equations of motion.
In particular, when changing variables from $(x,y)$ to $(u,v)$,
the Laplacian retains its form
$\partial_x^2+\partial_x^2\propto \partial_u^2+\partial_v^2$.

An alternative interpretation of the above separation of variables is as follows.
Given an arbitrary configuration for $\hat{n}$, consider the contour lines of
constant $\hat{n}_z$, the $\hat{z}$ component of the magnetization.
For contour
lines with $\hat{n}_z\ne \pm 1$ that form closed curves, consider the winding
number of $\hat{n}_x+i\hat{n}_y=\sin(\alpha)e^{i\beta}$.
For smooth configurations, 
this is a quantized integer that cannot change between 
neighboring contours which
do not cross $\hat{n}_z\ne \pm 1$.  This implies the winding number is constant
in regions between contours with $\hat{n}_z\ne \pm 1$.
Label different contours of $\hat{n}_z$ by $u$ and
the label coordinate along each contour by $v$.

\begin{figure}
\begin{center}
\begin{tabular}{cc}
\includegraphics[width=1.5in]{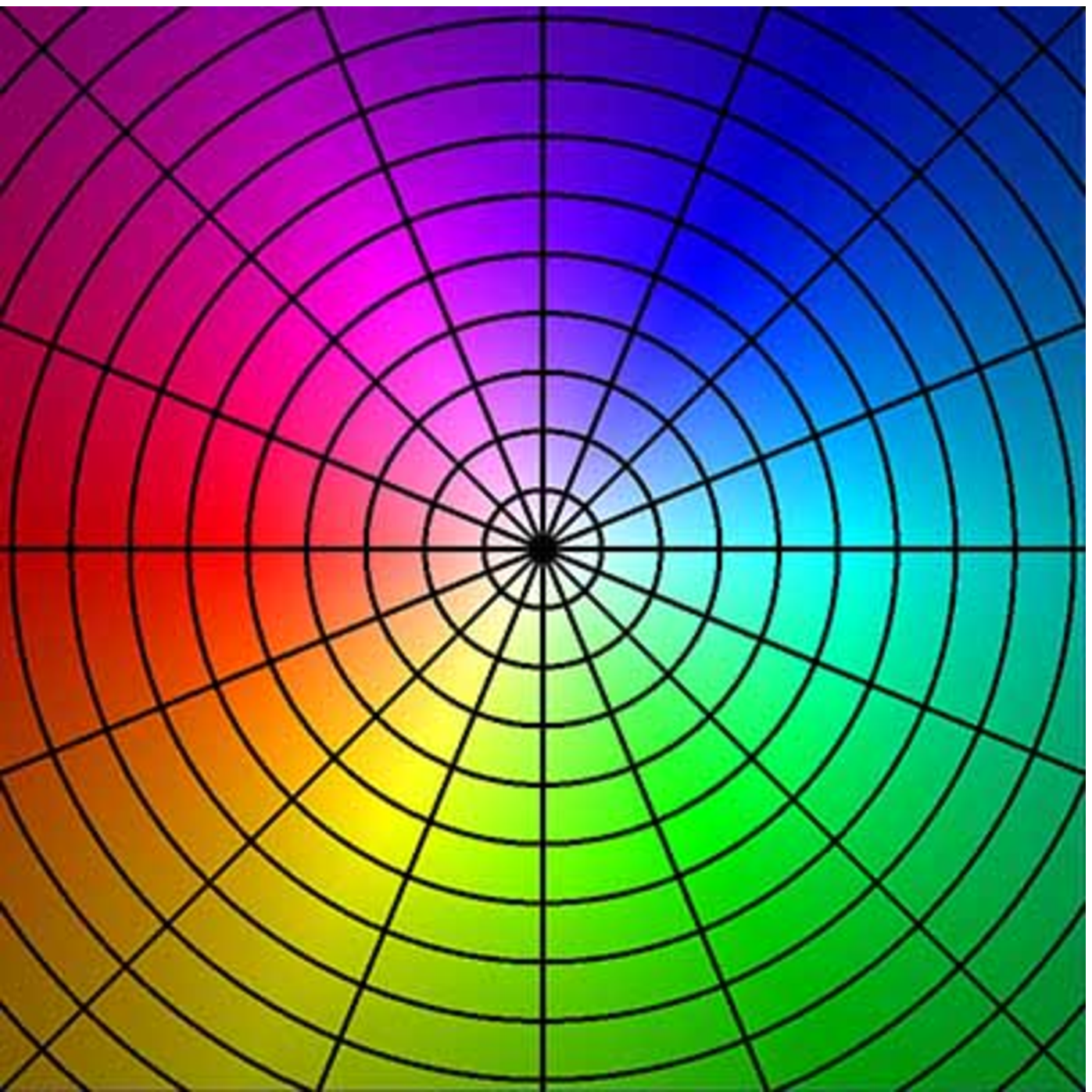}
\includegraphics[width=1.5in]{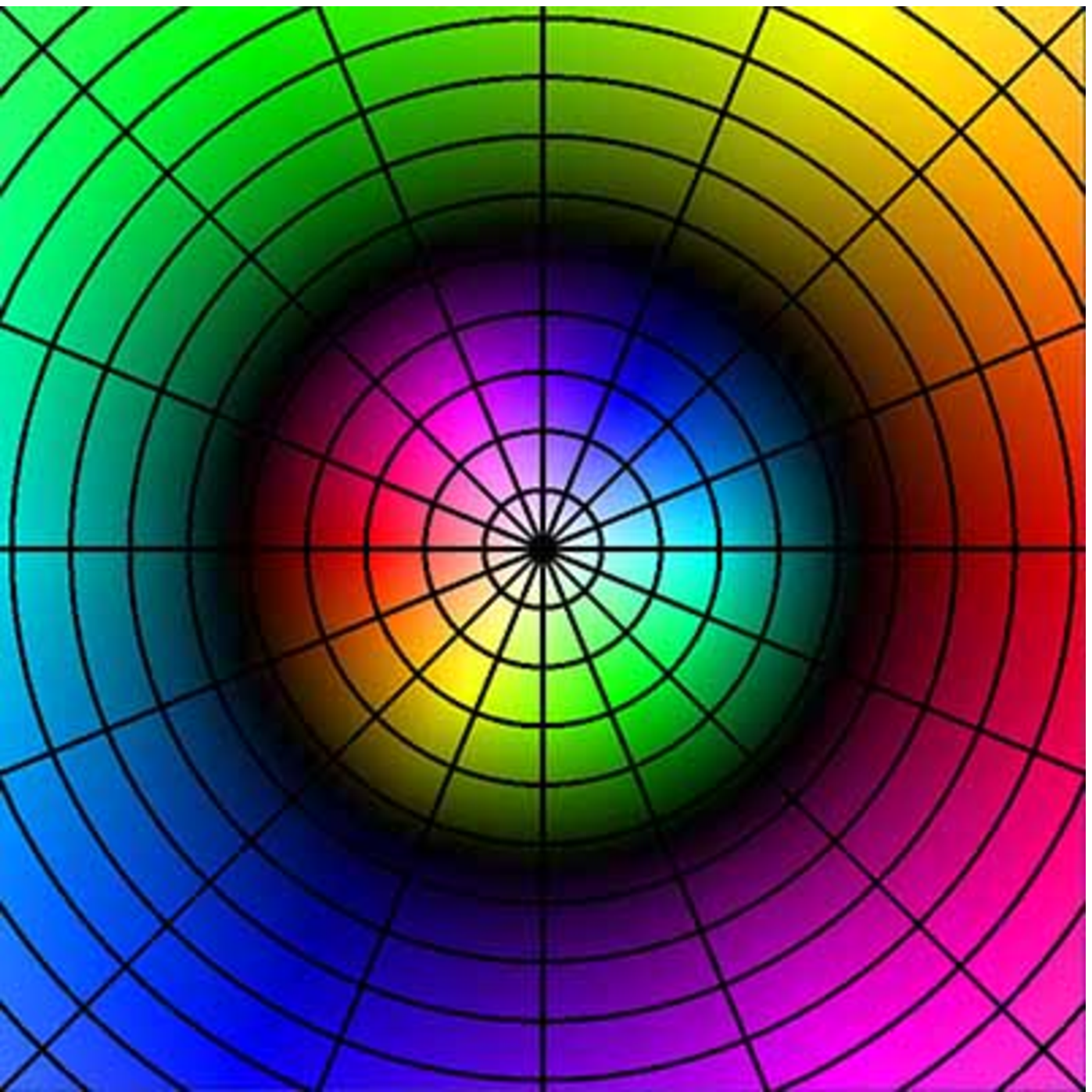}
\end{tabular}
\end{center}
\caption[Two dimensional skyrmion and anti-skyrmion solutions versus contour lines]{
For an arbitrary smooth magnetization configuration, contour lines of $\hat{n}_z$
and the phase of $\hat{n}_x+i\hat{n}_y$ provide a natural coordinate system.
For skyrmion solutions of the ordinary ferromanget, this gives an orthogonal
coordinate system with contour lines intersecting at right angles.
Using the same ansatz for spinor condensate ferromagnets where contour
lines intersect at right angles allows us to solve the non-linear and non-local
equations of motion.
The magnetization and contour lines are shown for the single skyrmion 
(left) for the ordinary ferromagnet and 
neutral configuration (right) for the spinor condensate ferromagnet.
Hue indicates orientation of $\hat{n}_x$, $\hat{n}_y$ 
components of the magnetization and 
brightness gives the
$\hat{n}_z$ component with white (black) indicating 
$\hat{n}_z=+1$ ($\hat{n}_z=-1$).
}
\label{chap:skyrmion:fig:contours}
\end{figure}

In order to have a non-zero winding number, $\beta$ must have some dependence
on $v$ and the minimal one is $\beta\propto v$.  
In principal, $\beta$ can also depend on $u$ and have some
non-monotonic dependence on $v$, but linear dependence is
the smoothest one compatible with non-zero winding number.
We see that skyrmion solutions can be interpreted in the above manner along
with the additional condition that $u$ and $v$ are mutually othogonal
and satisfy Laplace's equation.  Physically, these additional conditions
on $u$ and $v$ can be understood as a consequence of minimizing the spin
stiffness.  
The relationship between contour lines, winding number,
and the exact solutions we discussed in this section 
is shown in Fig. \ref{chap:skyrmion:fig:contours}.

With this viewpoint, we can now generalize the skyrmion solutions 
for ordinary ferromagnets and also find new solutions for spinor
condensate ferromagnets.  
For $f(x+iy)=u(x,y)+iv(x,y)$ holomorphic, 
we take 
\begin{align}
\alpha&=\alpha(u),& \beta=kv
\label{chap:skyrmion:eq:general_solution}
\end{align} 
for the parameterization of Eq. \ref{chap:skyrmion:eq:skyrmion}.
Compared to the skyrmion solutions of Eq. \ref{chap:skyrmion:eq:skyrmion_solution} 
with $\tan(\alpha/2)=\exp[u]$ for ordinary ferromagnets, 
we allow for general dependence $\alpha(u)$ for spinor ferromagnets.
Whereas for ordinary ferromagnets we only need to solve $\nabla^2 \hat{n}=0$,
for spinor condensate ferromagnets we need to solve Eqs.
\ref{chap:skyrmion:eq:eom}, \ref{chap:skyrmion:eq:superfluid} with $\partial_t=0$.

In addition, we include a constant of proportionality $\beta=kv$ instead
of $\beta=v$.  Recall  for ordinary ferromagnets,
$\beta=+v$ and $f(x+iy)$ holomorphic give skyrmion
solutions with positive skyrmion density $q$ 
while $\beta=-v$ and $f(x-iy)$ antiholomorphic
give anti-skyrmion solutions with $q$.
We can treat both types of solutions with just
$f(x+iy)$ holomorphic by allowing $\beta=kv$
with $k$ positive or negative.

There are two cases to consider for $f(z)$.
The first is when $f(z)$ is a polynomial in $z$ with no singularities.
This will turn out to describe skyrmion and anti-skyrmion stripe and domain wall
configurations for both ordinary and spinor condensate
ferromagnets.  
The second case is when $f(z)$ has singularities.
Since  $\hat{n}$ should be single-valued,
$\beta$ and thus $kv$ can only have constant $2\pi N$ 
discontinuities 
with $N$ integer.
This implies $f(z)$ can only have logarithmic singularities.
These solutions will turn out to simply be the localized skyrmion configurations
for ordinary ferromagnets and neutral collections of localized 
skyrmions and anti-skyrmions for spinor condensate ferromagnets.

Next we consider Eq. \ref{chap:skyrmion:eq:superfluid} for the superfluid velocity 
$\mathbf{v}$.  For the parameterization in Eqs. \ref{chap:skyrmion:eq:skyrmion},
\ref{chap:skyrmion:eq:general_solution}, we see from Eq. \ref{chap:skyrmion:eq:skyrmion_density}
that the skyrmion density $q$ only depends on $u$.
We thus take $\Phi(u)$ to only depend on $u$
which reduces the equation $-\nabla^2\Phi=q$ to 
$-\Phi''(u)=q(u)$.
From here on primes denote derivatives with respect to $u$.
By solving for $\Phi(u)$ we can then obtain $\mathbf{v}$ by differentiating.
Explicitly, we obtain 
\begin{align}
q&=-k\cos(\alpha)'|\partial_z f|^2,&
\mathbf{v}_z &=iSk[C+\cos(\alpha)]\partial_z f
\label{chap:skyrmion:eq:q_v}
\end{align}
where $\mathbf{v}_z=\mathbf{v}_x-i\mathbf{v}_y$,
$\partial_z f=\partial_x f-i\partial_y f$ 
and $C$ is a constant of integration physically describing a
$u$ independent constant contribution to the superfluid velocity.

We now proceed to analyze Eq. \ref{chap:skyrmion:eq:eom} for spinor condensate ferromagnets.
By substituting the results of Eq. \ref{chap:skyrmion:eq:q_v} above and the 
parameterization in Eqs. \ref{chap:skyrmion:eq:skyrmion}, \ref{chap:skyrmion:eq:general_solution},
we find the $\hat{z}$ component of Eq. \ref{chap:skyrmion:eq:eom} is automatically satisfied.
In addition, the $\hat{x}$ and $\hat{y}$ components are proportional to 
each other and
reduce to a second ordinary differential equation for $\alpha(u)$.
For completeness, we can use the same approach to analyze 
Eq. \ref{chap:skyrmion:eq:ferromagnet_lagrangian} for ordinary ferromagnets using the
same parameterization in Eqs. \ref{chap:skyrmion:eq:skyrmion}, \ref{chap:skyrmion:eq:skyrmion_solution}.

For ordinary and spinor condensate ferromagnets, the equations of motion in 
Eq. \ref{chap:skyrmion:eq:ferromagnet_lagrangian} and Eq. \ref{chap:skyrmion:eq:eom} reduce to
\begin{align}
\nonumber
2\alpha''&=+k^2\sin(2\alpha)\\
2\alpha''&=-4SCk^2\sin(\alpha)-(2S-1)k^2\sin(2\alpha)
\label{chap:skyrmion:eq:eom_reduced}
\end{align}
respectively.  Notice the equations of motion
for ordinary ferromagnets are formally
given by the $S=0$ limit for spinor condensate ferromagnets.
Recall the spin stiffness term scales linearly with $S$
whereas the superfluid kinetic energy scales quadratically with $S$.
For spinor condensate ferromagnets in the limit $S\rightarrow 0$, 
the superfluid kinetic energy is negligible compared to the spin stiffness 
and the ordinary ferromagnet is recovered.  From here on, we consider
the more general equation of motion for spinor condensate ferromagnets.

Interpreting $u$ as time, this equation
is that of a classical particle with coordinate $\alpha$
and momentum $\alpha'$.  The total energy $E=K+U$
is a constant of motion where $K=(\alpha')^2/2$ is the kinetic
energy while the periodic potential $U$ and equation of motion
are
\begin{align}
\nonumber
U(\alpha)&=-2SCk^2\cos(\alpha)-\frac{2S-1}{4}k^2\cos(2\alpha)\\
\alpha'&=\sqrt{E-U(\alpha)}
\label{chap:skyrmion:eq:potential}
\end{align}
with $S=0$ for ordinary ferromagnets.
For this type of solution, 
the total energy calculated from the
Hamiltonian in Eq. \ref{chap:skyrmion:eq:final_lagrangian} is given by 
\begin{align}
\mathcal{H}&=N\int dudv\frac{S}{2}
\left[\frac{2S+1+4C^2S}{4}k^2-E+\alpha'^2\right]
\label{chap:skyrmion:eq:hamiltonian_reduced}
\end{align}
and similarly, $S=0$ for the term in brackets for
ordinary ferromagnets.

The integer $N$ comes from changing variables
$(x,y)$ to $(u,v)$ and taking into account each 
$(u,v)$ may occur for multiple $(x,y)$.  
Mathematically, it is given by the degree of $f$
viewed as a map from the complex plane to itself
$\mathbb{C}\rightarrow\mathbb{C}$.
For localized skyrmion solutions for the ordinary ferromagnet, it
physically corresponds to the skyrmion number.
As an example, $f(z)=\log z$ for
the solutions shown in Fig. \ref{chap:skyrmion:fig:skyrmion_single}
(see Eq. \ref{chap:skyrmion:eq:localized_sphere} for a generalization)
and each value of $(u,v)$ occurs exactly once for $(x,y)$ 
ranging over the plane and thus $N=1$.
For $f(z)=\log[
(\vartheta(z-\lambda,i)\vartheta(z-\lambda^*,i))/
(\vartheta(z+\lambda,i)\vartheta(z+\lambda^*,i))]$ 
with $\vartheta(z,\tau)$ the elliptic theta function
and $\lambda=(1+i)/2$
for the solutions shown in 
Fig. \ref{chap:skyrmion:fig:skyrmion_lattice} 
(see Eq. \ref{chap:skyrmion:eq:localized_torus} for a generalization)
and each value of $(u,v)$ occurs
exactly twice for $(x,y)$ ranging over the unit cell and thus $N=2$.

Here we comment on the significance of the parameters $C$ and $E$.
Different values of $C$ and $E$ correspond to solutions with 
\textit{different boundary conditions}.  Physically, $C$ controls 
a constant background contribution to the superfluid velocity.
The parameter $E$ controls the relative scaling of the two components
of the coordinate system.  For example, for doubly periodic 
stripe solutions which we will discuss later, $E$ controls the aspect ratio
of the unit cell.

Formally, $E$ and $C$ are constants of integration for the equations of motion.
Since Eq. \ref{chap:skyrmion:eq:eom_reduced} is a second order differential
equation, we have to specify both $\alpha(u)$ and 
$\alpha'(u)$, with the latter given indirectly by the 
constant of motion $E$.
In addition, $C$ enters through integration of Eq. \ref{chap:skyrmion:eq:q_v}
relating the skyrmion density to the superfluid velocity.

These considerations
mean that the total energy in Eq. \ref{chap:skyrmion:eq:hamiltonian_reduced}
cannot be directly compared for different $C$ and $E$.
In particular, one should not consider minimizing the total energy
$\mathcal{H}$ with respect to $E$ and $C$.
Specific values of $E$, and $C$ will be selected by terms beyond the 
non-linear sigma model considered in this paper.

\subsection{Localized skyrmions and anti-skyrmions}
\label{subchap:skyrmion:sec:localized}

We begin by considering the case of localized skyrmions for ordinary
ferromagnets and neutral collections of localized skyrmions and
anti-skyrmions for spinor condensate ferromagnets.
This corresponds to $f(z)$ having logarithmic singularities.
The singularities should be of integer magnitude and $k$ should
also be an integer.

Requiring a well-behaved, finite energy solution gives rise to several
constraints.
Consider Eq. \ref{chap:skyrmion:eq:q_v} for the skyrmion density $q$
and supefluid velocity $\mathbf{v}$.
Since $\partial_z f$ diverges at the logarithmic singularities,
we require $\cos(\alpha)'$ and $C+\cos(\alpha)$ to vanish.
In addition, a finite region in $(x,y)$ near logarithmic 
singularities is mapped to an infinite region in $(u,v)$.
The constant term in Eq. \ref{chap:skyrmion:eq:hamiltonian_reduced} 
for the total energy is then integrated 
over an infinite interval.
Thus, we also require it to vanish.

These constraints uniquely specify $E$, $C$ and the asymptotic 
value $\alpha_{-\infty}$.  For spinor condensate ferromagnets,
$E=k^2(1+6S)/4$, $C=\pm 1$, $\cos(\alpha_{-\infty})=\mp 1$ with 
$S=0$ for ordinary ferromagnets.
For $C=\pm 1$, we find for spinor condensate ferromagnets
the solution of Eq. \ref{chap:skyrmion:eq:eom_reduced} and the total energy
of Eq. \ref{chap:skyrmion:eq:hamiltonian_reduced} given by
\begin{align}
\nonumber
\alpha(u)&=\cos^{-1}(\mp 1)-2\cot^{-1}(\sqrt{2S}\sinh(ku))\\
\mathcal{H}&=4\pi N S k^2
\left(1+\frac{2S\tan^{-1}(\sqrt{2S-1})}{\sqrt{2S-1}}\right)
\label{chap:skyrmion:eq:sol_neutral}
\end{align}
which describes either a $0\rightarrow2\pi$ or
$-\pi\rightarrow+\pi$ kink solution.
For $C=\pm 1$ we find for ordinary ferromagnets
\begin{align}
\nonumber
\alpha(u)&=\cos^{-1}(\mp 1)+2\tan^{-1}(e^{ku})\\
\mathcal{H}&=2\pi N S k^2
\label{chap:skyrmion:eq:sol_charged}
\end{align}
which describes in contrast either a $0\rightarrow\pi$
or $-\pi\rightarrow0$ kink solution.

Consider the classical mechanics problem describing the evolution of $\alpha(u)$
in Eq. \ref{chap:skyrmion:eq:potential}.   For these solutions, $E$ lies at the maximum
giving rise to kink solutions.  For ordinary ferromagnets, the kinks
connect $0\rightarrow\pi$ or $-\pi\rightarrow 0$ 
and carry net positive or negative skyrmion charge, respectively.
For spinor condensate ferromagnets, the kinks connect
$-\pi\rightarrow +\pi$ or $0\rightarrow 2\pi$ 
and carry net neutral skyrmion charge.
The neutral configurations consist of regions of oppositely 
charged skyrmion and anti-skyrmions.
These regions are separated by lines where the skyrmion density 
$q$ vanishes, the magnetization $\hat{n}$ is along $\hat{z}$, 
and the superfluid velocity $\mathbf{v}$ is large. 

For $f(z)$ having a finite number of logarithmic singularities, we can write
\begin{align}
f(z)=\log\left[f_0\frac{\prod_{n=1}^{N_a}(z-a_n)}{\prod_{m=1}^{N_b} (z-b_m)}\right]
\label{chap:skyrmion:eq:localized_sphere}
\end{align}
with the degree $N$ of the function $f(z)$ given by $N=\text{max}(N_a,N_b)$.
Here, $a_n$ ($b_n$) give the locations of $+\hat{z}$ ($-\hat{z}$) merons
each carrying net skyrmion charge $2\pi$
for the ordinary ferromagnet.  
In contrast,  $a_n$ ($b_n$) give the locations
of skyrmions (anti-skyrmions) each carrying net skyrmion charge $+4\pi$ ($-4\pi)$
for the spinor condensate ferromagnet.
We show the corresponding plots of $\hat{n}$, $q$, and $\mathbf{v}$ in 
Fig. \ref{chap:skyrmion:fig:skyrmion_single} for $f(z)=\log(z)$.
The single skyrmion solution for the ordinary ferromagnet with $S=0$ is
shown on the left and a neutral configuration of one skyrmion
and one anti-skyrmion for the spinor condensate ferromagnet with $S=1$
is shown on the right.

For a periodic lattice of logarithmic singularities,
\begin{align}
f(z)=\log\left[f_0\frac{\prod_{n=1}^{N_a}\vartheta(z-a_n,\tau)}
{\prod_{m=1}^{N_b} \vartheta(z-b_m,\tau)}\right]
\label{chap:skyrmion:eq:localized_torus}
\end{align}
where $f_0$ is a constant, and $1$, $\tau$ give the basis vectors generating
the latttice in complex form,
and $\vartheta(z,\tau)$ is the elliptic theta function.
The elliptic theta function $\vartheta(z,\tau)$ is essentially
uniquely specified by the quasiperiodic condition
\begin{align}
\vartheta(z+n+m\tau,\tau)=\exp(-\pi i m^2 \tau-2\pi i m z)\vartheta(z,\tau)
\label{chap:skyrmion:eq:theta}
\end{align}
and holomorphicity.  Just as Eq. 
\ref{chap:skyrmion:eq:localized_sphere} is built up from 
the linear polynomials $(z-z_0)$ which are holomorphic and vanish at one point
in the complex plane, Eq. \ref{chap:skyrmion:eq:localized_torus}
is built up from $\vartheta(z,\tau)$ which are holomorphic and vanish at one point
in the unit cell.  For a discussion of theta functions in the quantum
Hall effect, see Ref. \cite{haldane-85}.

In the lattice case,
$N_a=N_b$ and $\sum a_n=\sum b_n$ in order to have $f(z)$ periodic.
This restriction comes from requiring $f(z+n+m\tau)=f(z)$ and using 
Eq. \ref{chap:skyrmion:eq:theta}.
The degree $N$ of the function $f(z)$ per unit cell is given by $N=N_a=N_b$.
Again, $a_n$, $b_n$ give the locations of merons (skyrmions or anti-skyrmions)
for the ordinary (spinor condensate) ferromagnet.
Fig. \ref{chap:skyrmion:fig:skyrmion_lattice} shows plots for $f(z)$ having a lattice of 
logarithmic singularities with $N=N_a=N_b=2$, $f_0=1$, $\tau=i$, $a_1=-a_2=(1+i)/2$, 
$b_1=-b_2=(1-i)/2$.  Again the ordinary (spinor condensate) 
ferromagnet is on the left (right).

\subsection{Stripe configurations}

\label{subchap:skyrmion:sec:stripe}

Now we turn to case of stripe configurations described by $f(z)$ polynomial
in $z$.  The behavior of $\alpha(u)$ solutions
controlled by the potential in Eq. \ref{chap:skyrmion:eq:potential} changes as $E$ 
crosses critical points  $dU/d\alpha=0$ of the potential.

\begin{figure}
\begin{center}
\includegraphics[width=2.5in]{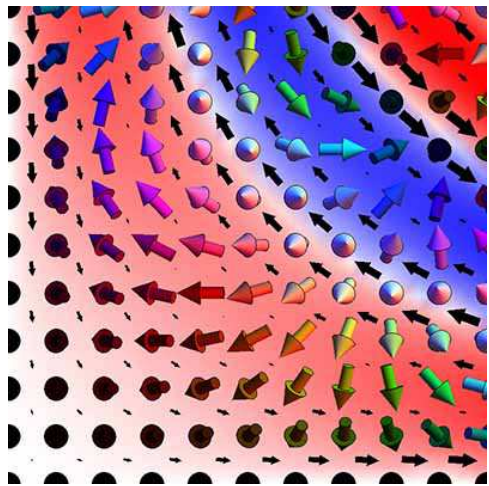}
\end{center}
\caption[Skyrmion and anti-skyrmion solution with corner boundary conditions]{
Neutral stripe configuration satisfying non-trivial corner boundary conditions.
Red (blue) background indicates positive (negative) skyrmion 
density $q$, black 2D arrows the superfluid velocity $\mathbf{v}$, 
and shaded 3D arrows the magnetization $\hat{n}$.
}
\label{chap:skyrmion:fig:skyrmion_corner}
\end{figure}

For completeness, we first briefly consider $f(z)$ given by higher order 
polynomials.  The corresponding
stripe solutions are not doubly periodic, but satisfy non-trivial
boundary conditions.  For example, we show the magnetization $\hat{n}$, 
skyrmion density $q$, and superfluid velocity $\mathbf{v}$ for $f(z)=iz^2$
in Fig. \ref{chap:skyrmion:fig:skyrmion_corner}.  This solution satisfies corner boundary
conditions with zero normal component to both the superfluid velocity $\mathbf{v}$
and spin current 
$\mathbf{J}^i_\mu=\hat{n}^i\mathbf{v}_{\mu}
-\epsilon^{ijk}\hat{n}^j\nabla_\mu\hat{n}^k/2$.

\begin{figure*}
\begin{center}
\includegraphics[width=5.75in]{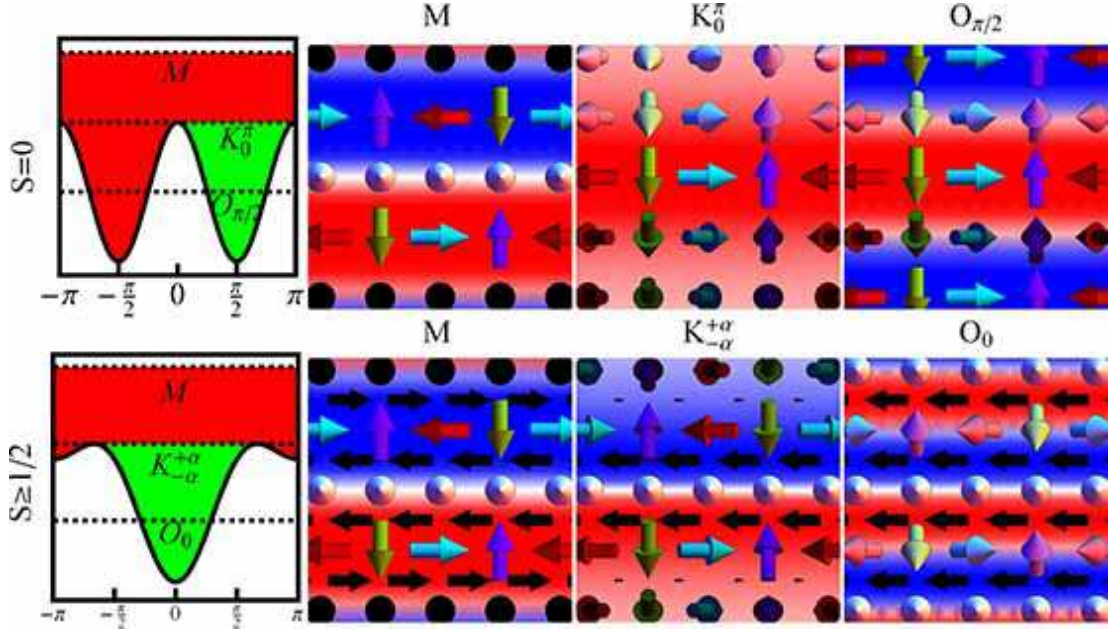}
\end{center}
\caption[Schematics for stripe solutions]{
Stripe configurations for different boundary conditions in
$S\ge1/2$ spinor condensate (bottom row) and
$S=0$ ordinary ferromagnets (top row).
Red (blue) background indicates positive (negative) skyrmion 
density $q$, black 2D arrows the superfluid velocity $\mathbf{v}$, 
and shaded 3D arrows the magnetization $\hat{n}$.
Notice $\hat{n}_x$, $\hat{n}_y$ $(\hat{n}_z)$ 
wind horizontally (oscillate vertically).
The first column shows schematic plots of the classical 
periodic potential 
$U(\alpha)$ controlling the evolution of the angular variable 
$\alpha$ for $\hat{n}_z=\cos(\alpha)$.
Labels for the corresonding type of solution are below the
dashed lines indicating the corresponding constant of motion $E$.

For $E$ above the maximum of $U$, $\alpha(u)$ is montonic in 
the coordinate
$u$ giving rise to periodic stripe configurations labeled $M$ 
with $\hat{n}_z$ covering the entire range $\pm1$.
For $E$ at a potential maximum, kink solutions connecting the 
maxima $\alpha_1$, $\alpha_2$ give rise to single domain 
wall configurations in $\hat{n}_z$ labeled $K_{\alpha_1}^{\alpha_2}$.
For a given value of $C$, kink solutions occur at one specific value of $E$.
For $E$ below the maximum, $\alpha$ oscillates about a minimum located
at $\alpha_0$ giving rise to periodic stripe configuration with 
$\hat{n}_z$ oscillating about $\cos(\alpha_0)$.

Notice that monotonic solutions $M$ are qualitatively similar for both $S\ge 1/2$ spinor condensate and $S=0$ ordinary ferromagnets.
In contrast, spinor condensate (ordinary) ferromagnets have a 
single (two distinct)
$2\pi$ ($\pi$) kink solutions $K$ centered about 
$\alpha=0$ ($\alpha=\pm\pi/2$).
In addition, there is a single 
(two distinct) oscillatory solutions $O$ 
also centered about $\alpha=0$ ($\alpha=\pm\pi/2$) 
for spinor condensate 
(ordinary) ferromagnets.
}
\label{chap:skyrmion:fig:stripes}
\end{figure*}

From here on, we focus on $f(z)=iz$ with the corresponding solutions doubly periodic 
and describing stripe configurations.
The different types of behavior for $\alpha(u)$ are illustrated schematically
along with the resulting configurations for $\hat{n}$ in Fig. \ref{chap:skyrmion:fig:stripes}.
$E$ above the global maximum corresponds to $\alpha(u)$ monotonic in $u$
which from here on we denote as $M$.  This solution describes
a periodic stripe solution with $\hat{n}_z$ varying over the entire
range $\pm1$.
$E$ at a local maximum corresponds to a kink solution for $\alpha(u)$
connecting $\alpha_1$ to $\alpha_2$ denoted as $K_{\alpha_1}^{\alpha_2}$.
This solution describes a single domain wall configuration
in $\hat{n}_z$ and is the analog of the localized solution described earlier.
$E$ below a local maximum corresponds to $\alpha(u)$
oscillating near a fixed value $\alpha_0$
denoted as $O_{\alpha_0}$.  Finally, $E$ below the global minimum 
is forbidden, denoted as $F$.

Notice that for $S=0$, $S=1/2$, $S>1/2$,
the $\cos(2\alpha)$ term in the potential $U(\alpha)$
of Eq. \ref{chap:skyrmion:eq:potential} is negative, zero, and positive.
For the montonic solutions $M$, this does not affect the qualitative behavior
of the resulting periodic stripe configurations.
For kink solutions $K$ connecting
$0\rightarrow 2\pi$ or $-\pi\rightarrow +\pi$
($0\rightarrow \pi$ or $-\pi\rightarrow 0$), 
the resulting single domain wall carrys zero net
skyrmion charge (positive or negative skyrmion charge) for spinor 
condensate (ordinary) ferromagnets.
Oscillatory solutions $O$ also have different behavior with oscillations
centered about
$\hat{n}_z\approx \pm 1$ ($\hat{n}_z \approx 0$)
for spinor condensate (ordinary) ferromagnets.

\begin{figure*}
\begin{center}
\includegraphics[height=1.74in]{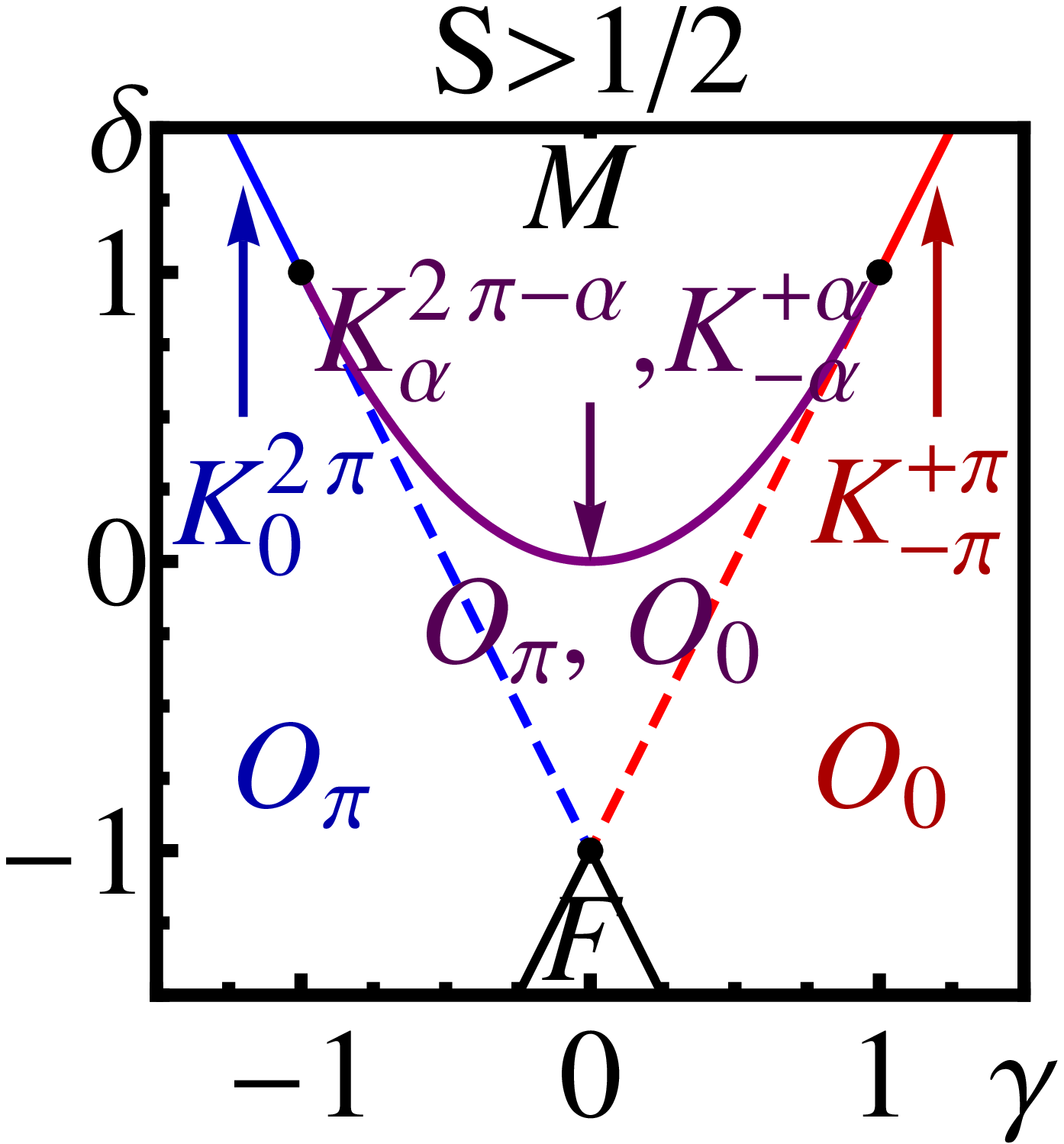}
\includegraphics[height=1.53in]{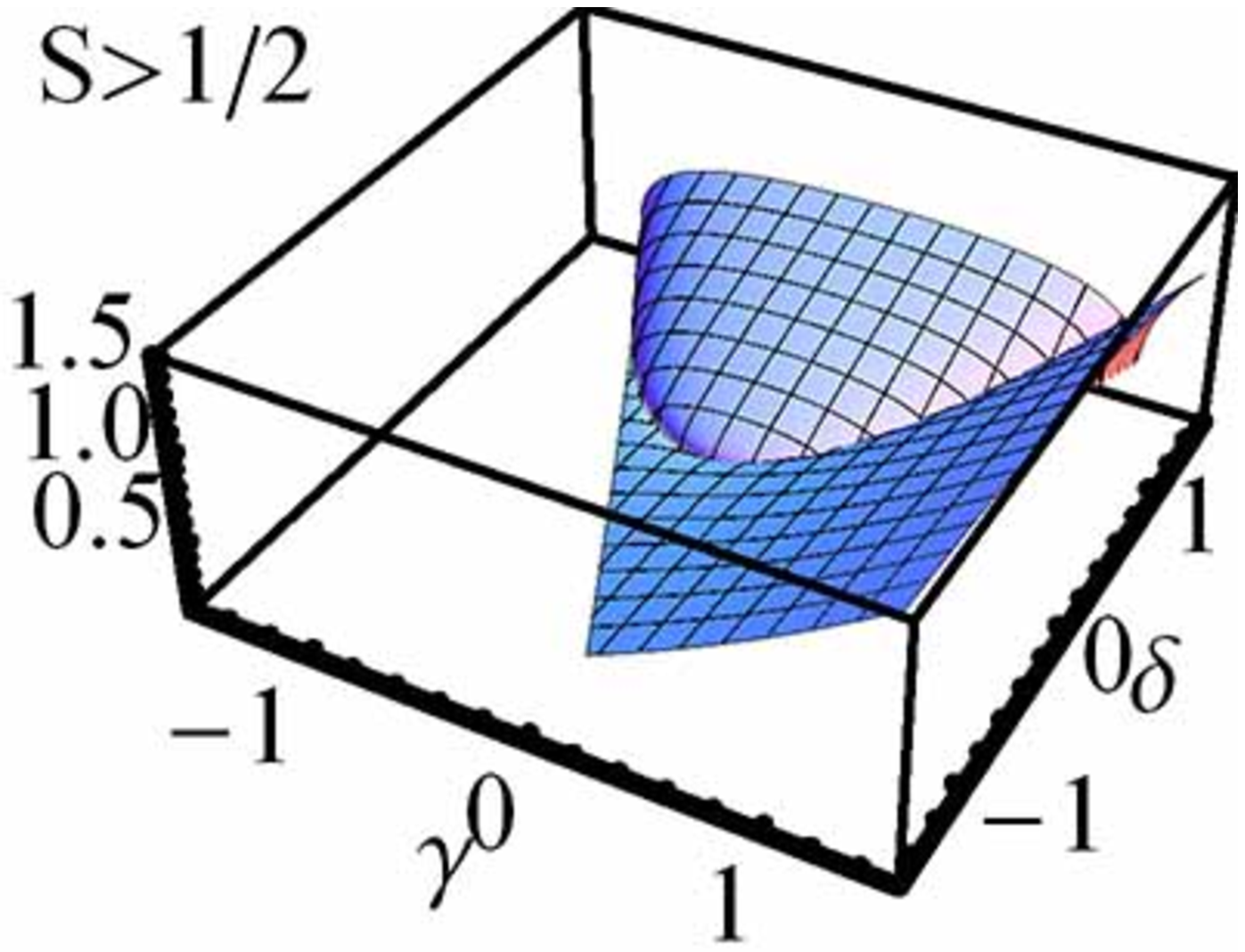}
\includegraphics[height=1.53in]{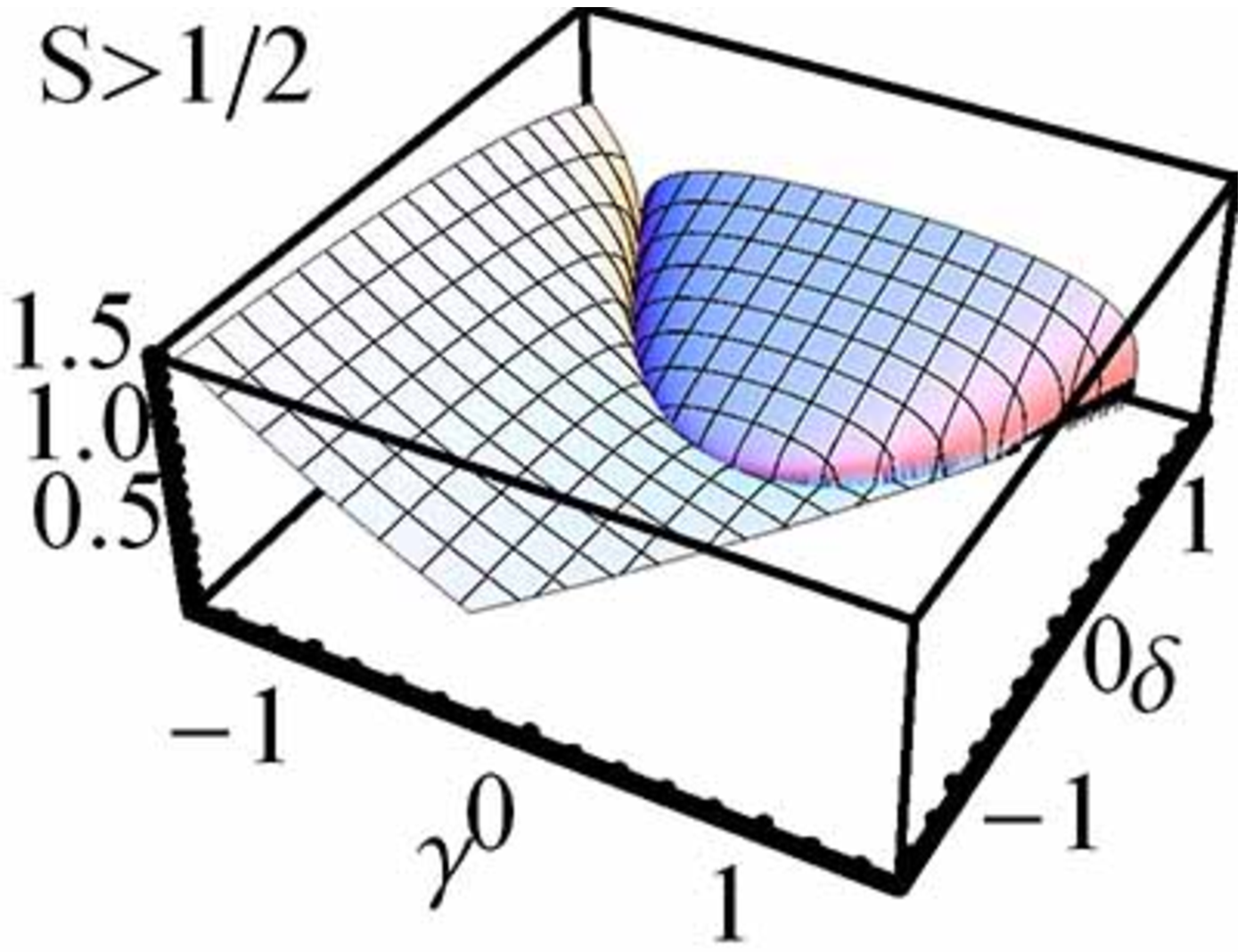}
\end{center}
\includegraphics[height=1.39in]{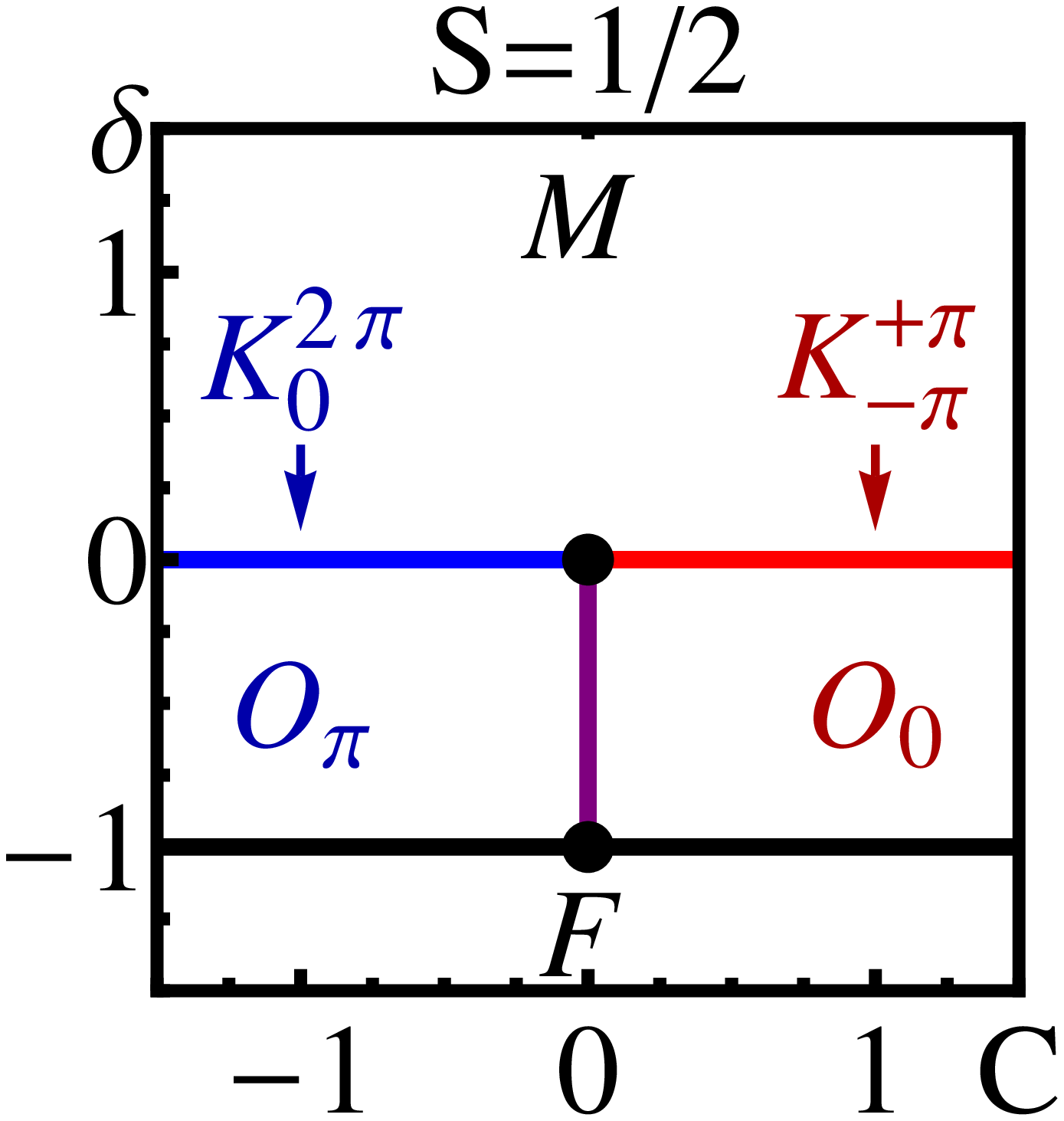}
\includegraphics[height=1.22in]{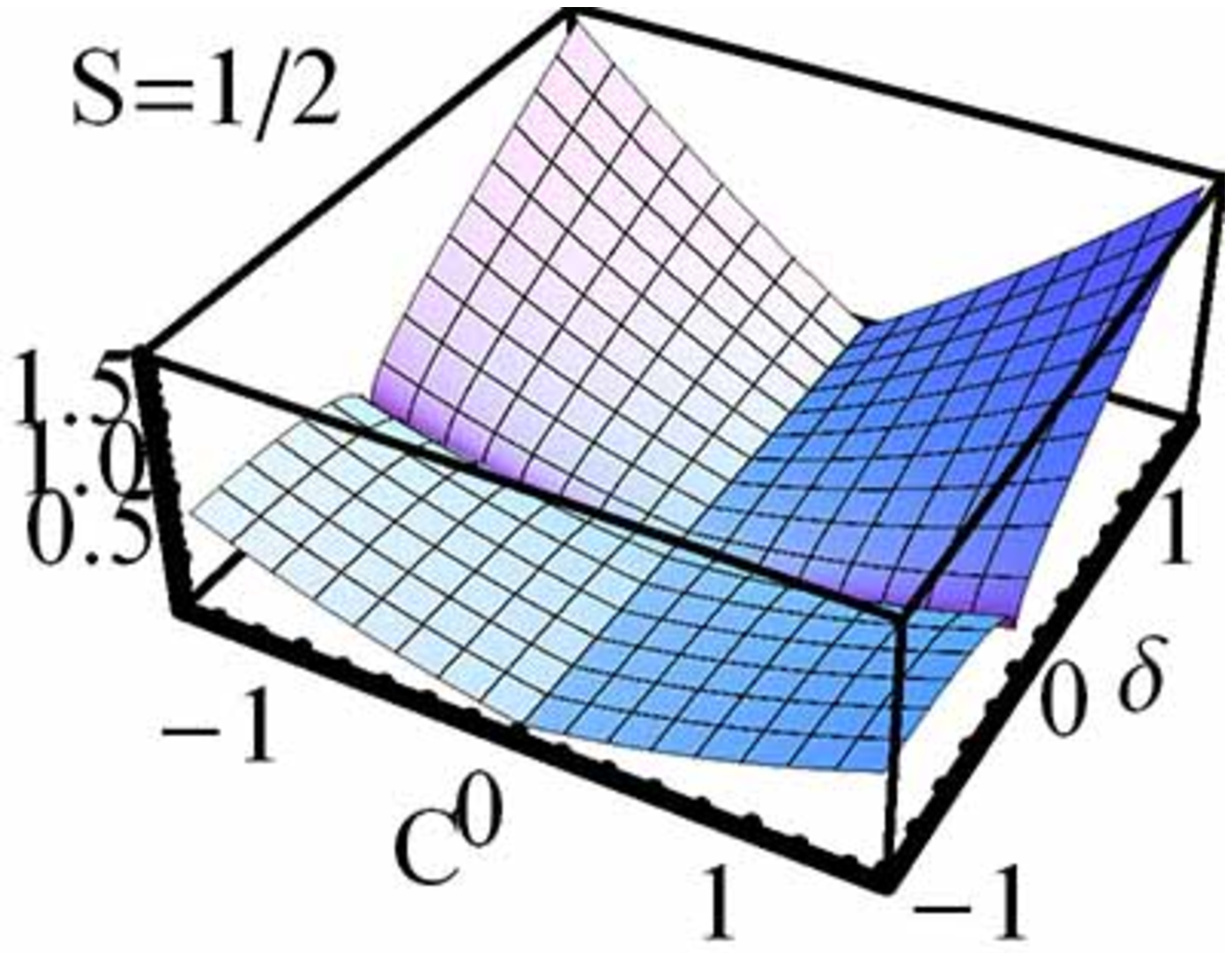}
\includegraphics[height=1.39in]{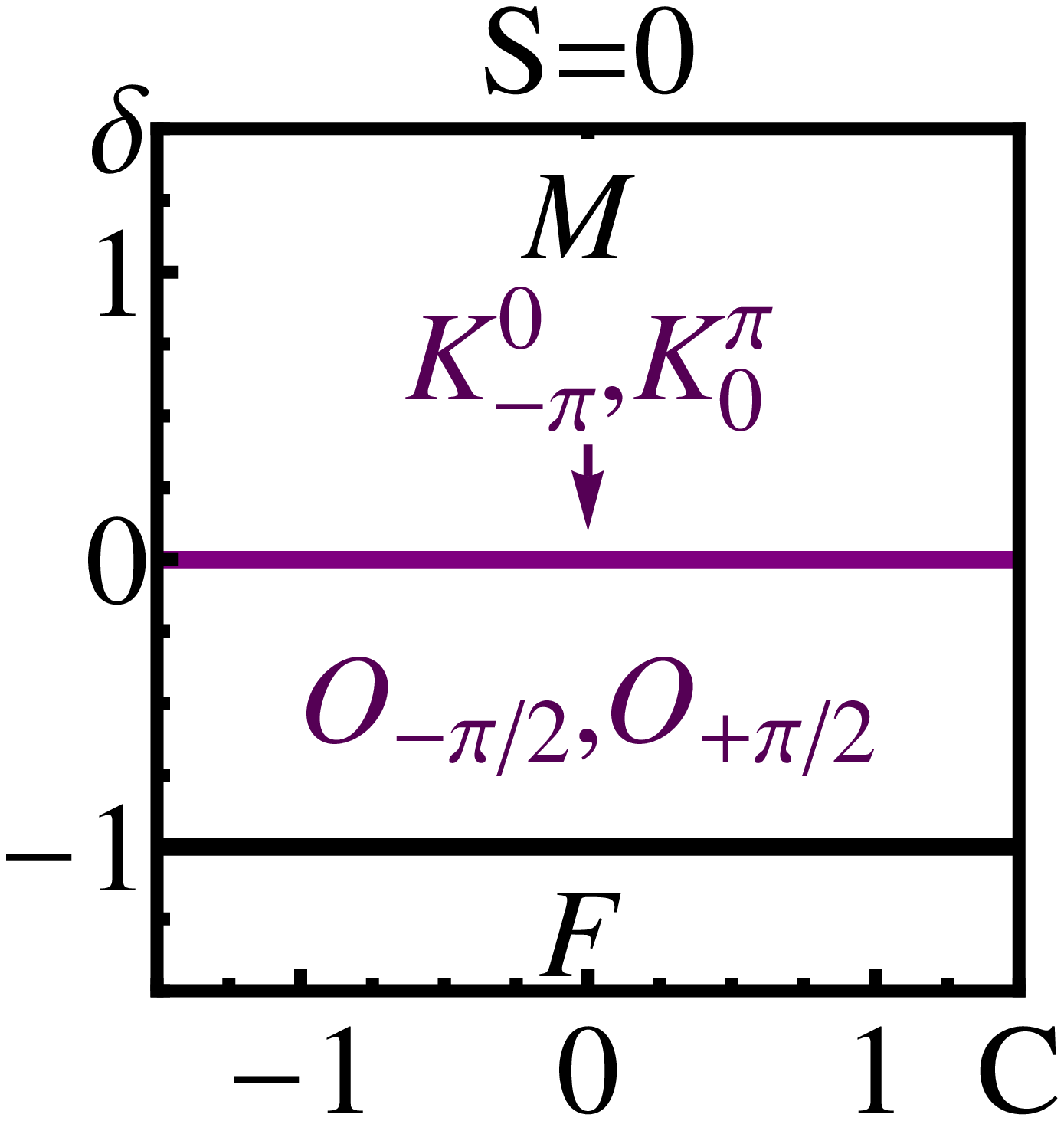}
\includegraphics[height=1.22in]{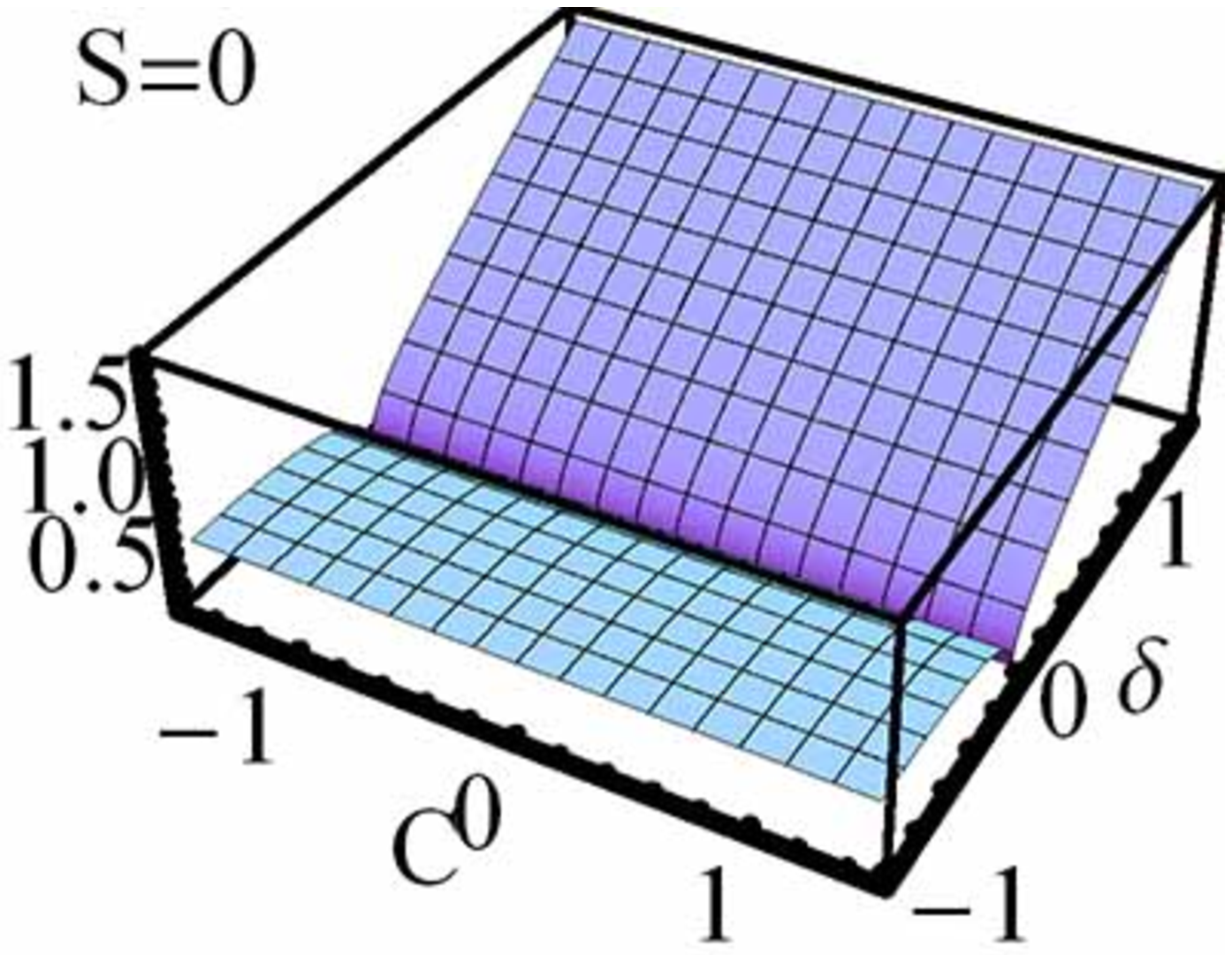}
\begin{center}
\end{center}
\caption[Classification of stripe solutions]{
Classification of solutions 
(2D plots) and total energy density (3D plots) for
stripe configurations given different
boundary conditions in $S>1/2$ (top row) and $S=1/2$ (bottom left)
spinor condensate as well as $S=0$ (bottom right)
ordinary ferromagnets.
Fig. \ref{chap:skyrmion:fig:stripes} illustrates the corresponding configurations.
$\gamma$ controls a constant contribution to the superfluid velocity.
$\delta$ controls the energy of an associated classical mechanics
problem giving the evolution of $\hat{n}_z=\cos(\alpha)$, the $\hat{z}$ 
component of the magnetization.
Monotonic solutions $M$ have $\hat{n}_z$ covering the entire range $\pm 1$.
Kink solutions $K_{\alpha_1}^{\alpha_2}$ describe single domain wall configurations
with $\hat{n}_z$ connecting $\cos(\alpha_1)$ to $\cos(\alpha_2)$.
Oscillatory solutions $O_{\alpha_0}$ have $\hat{n}_z$ oscillating
about  $\cos(\alpha_0)$.
Notice kink solutions $K$ always separate monotonic $M$ from oscillatory $O$
solutions.
For $S>1/2$ spinor condensate ferromagnets,
notice the two distinct oscillatory and kink solutions 
for each $\gamma$, $\delta$ in the region near the origin 
which are absent for $S=1/2$.
For $S=0$ ordinary ferromagnets, there is no dependence on $C$.
}
\label{chap:skyrmion:fig:phase_diagram}
\end{figure*}

For ordinary ferromagnet with $S=0$, we parametrize 
$E=k^2(1+2\delta)/4$ and find the solution of 
Eq. \ref{chap:skyrmion:eq:eom_reduced} and the total energy
of Eq. \ref{chap:skyrmion:eq:hamiltonian_reduced} given by
\begin{align}
\nonumber
\alpha(u)&=\pi/2\pm
\text{am}\left(ku\sqrt{1+\delta},(1+\delta)^{-1}\right)\\
\bar{\mathcal{H}}&=\frac{Sk^2}{2}
\left[-\frac{\delta}{2}+\theta(1+\delta)
\right]
\label{chap:skyrmion:eq:result_0}
\end{align}
where $\bar{\mathcal{H}}$ is the total energy density
given by the averaging $\mathcal{H}$ over the unit cell.
Also, $\text{am}(x,m)$ is the Jacobi amplitude function and we define
\begin{align}
\theta(m)=m\frac{\text{Re}[E(m^{-1})]}{\text{Re}[K(m^{-1})]}
\end{align}
with $K$ ($E$) the complete elliptic integral of 
the first (second) kind.
For $\delta<-1$ the solution is forbidden $F$.
For $-1\le\delta<0$ there are two oscillatory solutions
at the $\pm\pi/2$ minima $O_{\pm\pi/2}$.
For $\delta=0$ there are two kink solutions $K_0^{+\pi}$ and $K_{-\pi}^0$.
For $\delta>0$ the solution is monotonic $M$.
We show the classification of solutions along with the
total energy density for ordinary ferromagnets in the bottom left of 
Fig. \ref{chap:skyrmion:fig:phase_diagram}.  Notice the solutions
and total energy density do not depend on $C$.  This is because
$C$ enters through the superfluid velocity which is absent
from the Lagrangian for ordinary ferromagnets.

For spinor condensate ferromagnets with $S=1/2$, we parameterize the
constant of motion $E=|C|k^2(1+2\delta)$ and find the solution of 
Eq. \ref{chap:skyrmion:eq:eom_reduced} and the total energy density
from Eq. \ref{chap:skyrmion:eq:hamiltonian_reduced} given by
\begin{align}
\nonumber
\alpha(u)&=\cos^{-1}(C/|C|)+
2\text{am}\left(ku\sqrt{|C|(1+\delta)},(1+\delta)^{-1}\right)\\
\bar{\mathcal{H}}&=k^2
\left[\frac{(|C|-1)^2}{8}-\frac{\delta|C|}{2}+
|C|\theta(1+\delta)
\right]
\label{chap:skyrmion:eq:result_half}
\end{align}
depends only on $|C|$.
For $\delta<-1$ the solution is forbidden $F$.
For $-1\le\delta<0$ there is one oscillatory solution
$O_{\pi}$ ($O_{0}$) for $C<0$ ($C>0$).
For $\delta=0$ there is one kink solution
$K_0^{2\pi}$ ($K_{-\pi}^{+\pi}$) for $C<0$ ($C>0$).
For $\delta>0$ the solution is monotonic $M$.
We show the classification of solutions along with the
total energy density for $S=1/2$ spinor condensate ferromagnets in the 
bottom right of Fig. \ref{chap:skyrmion:fig:phase_diagram}.

For spinor condensate ferromagnets with $S>1/2$ we parameterize the
constant of motion $E=(2S-1)k^2(1+2\delta)/4$ and the constant 
$C=(2S-1)\gamma/2S$.  
With $\gamma=\tau$ and $\delta=-1+2\sigma \tau$ where $\sigma=\pm 1$
we find one solution for Eq. \ref{chap:skyrmion:eq:eom_reduced} 
while the total energy density from Eq. \ref{chap:skyrmion:eq:hamiltonian_reduced} given by 
\begin{align}
\nonumber
\alpha(u)&=
\text{Arg}\left[-\frac{\tau-\sigma+2\sqrt{\tau(\tau-\sigma)}s(u)+\tau s(u)^2}
{1+\sigma \tau-\sigma \tau s(u)^2}\right]\\
\bar{\mathcal{H}}&=\frac{S k^2}{2}\left[h_0+(2S-1)j_0\right]
\end{align}
where we define the auxiliary function
\begin{align}
s(u)&=\sin(ku\sqrt{(1-\sigma \tau)(2S-1)})
\end{align}
in the solution for $\alpha(u)$ and the functions
\begin{align}
\nonumber
h_0&=\frac{((2S-1)\tau-\sigma S)^2}{4S}\\
j_0&=\begin{cases}
\sigma \tau\sqrt{\tau-\sigma}&0\le\sigma \tau\le1\\
0&\text{otherwise}
\end{cases}
\end{align}
for the total energy density.
With $\gamma=r\sinh(\tau)$ and $\delta=-1+2r\sinh(\tau)$ we find two solutions
for Eq. \ref{chap:skyrmion:eq:eom_reduced} and the total energy density from
Eq. \ref{chap:skyrmion:eq:hamiltonian_reduced} given by
\begin{align}
\nonumber
\alpha_\pm(u)=&
\text{Arg}\left[-\frac{\cosh(\tau/4)e^{\pm i w(u)}\mp\sinh(\tau/4)}
{\sinh(\tau/4)e^{\pm i w(u)}\mp\cosh(\tau/4)}\right]\\
\bar{\mathcal{H}}=&\frac{S k^2}{2}\left[h+(2S-1)j\mp(2S-1)k\right]
\end{align}
where we define the auxiliary function
\begin{align}
w(u)=&\text{am}\left[ku\sqrt{2r(2S-1)},\frac{1+r-r\cosh(t)}{2r}\right]
\end{align}
in the solutions for $\alpha(u)$ and the functions
\begin{align}
\nonumber
h=&\frac{1-(2S-1)\delta}{2}+\frac{(2S-1)^2\beta^2}{4S}\\
\nonumber
j=&\frac{
\text{Re}[2\sqrt{r}E(\omega)]}
{\text{Re}[K(\omega)/\sqrt{r}]}+\\
\nonumber
&\frac{\text{Re}[2\sqrt{r}[\cosh(t/2)^2\Pi(-\sinh(t/2)^2,\omega)-K(\omega)]]}
{\text{Re}[K(\omega)/\sqrt{r}]}\\
\nonumber
k=&\begin{cases}
\frac{\text{Re}[\pi r\sinh(t)]}{\text{Re}[K(\omega)\sqrt{2/r}]},&
O_{\pi}, O_{0}\  \text{phase}\\
0&\text{otherwise}
\end{cases}\\
\omega=&\frac{1+r-r\cosh(\tau)}{2r}
\end{align}
for the total energy density 
where $\Pi(m,n)$ is the complete elliptic integral of the third kind.
We show the classification of solutions along with the
total energy density for $S>1/2$ spinor condensate ferromagnets in the 
top row of Fig. \ref{chap:skyrmion:fig:phase_diagram}.
The boundaries between solutions of different types are given by
$\delta=-1+2\beta$, $\delta=-1-2\beta$, $\delta=\beta^2$.
For increasing $\gamma$, 
notice kink solutions evolve from just one $K_{0}^{2\pi}$ through
a region with two $K_{\alpha}^{2\pi-\alpha}$ and $K_{-\alpha}^{+\alpha}$, 
to just one $K_{-\pi}^{+\pi}$ for $\gamma<-1$, $-1<\gamma<+1$, $+1<\gamma$,
respectively.  The kink solutions separate the monotonic solutions $M$
from the oscillatory solutions.  For increasing $\gamma$, 
the oscillatory solutions
also evolve from just one $O_\pi$ to a region with two
$O_\pi$ and $O_0$, to just one $O_0$ for $\delta<-1-2\gamma$,
$-1+2\gamma\ge\delta\ge-1-2\gamma$, $\delta<-1+2\gamma$, respectively.

\section{Discussion}
\label{chap:skyrmion:sec:discussion}

\begin{figure}
\begin{center}
\includegraphics[width=3in]{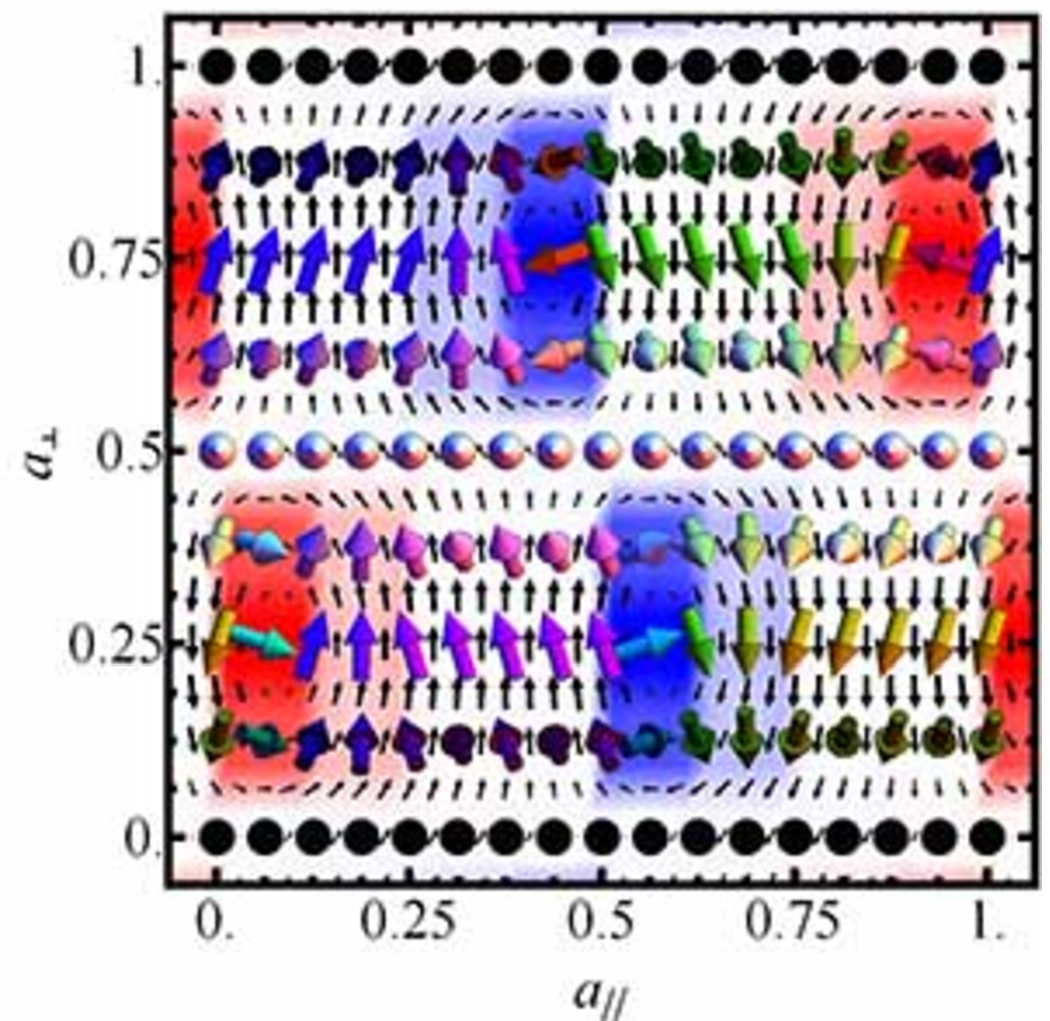}
\end{center}
\caption[Numerical solution with dipolar interactions]{
Numerically optimized configuration for two-dimensional 
spinor condensate ferromagnets with an effective
dipolar interaction modified by rapid Larmor precession. 
The magnetic field $\hat{B}=\hat{x}$ inducing Larmor precession 
lies along the horizontal axis in the plane.  
Lattice constants are $a_\parallel=90$ $\mu$m and $a_\perp=42$ $\mu$m.
Red (blue) background indicates positive (negative) skyrmion 
density $q$, black 2D arrows the superfluid velocity $\mathbf{v}$, 
and shaded 3D arrows the magnetization $\hat{n}$.
}
\label{chap:skyrmion:fig:dipolar}
\end{figure}

Having presented a unified description of both localized and extended stripe
solutions in ordinary and spinor condensate ferromagnets, we now turn
to how these solutions offer insight into different physical phenomena.
We first consider quantum Hall systems.  As discussed in Sec. \ref{chap:skyrmion:sec:qh},
configurations for quantum Hall ferromagnets away from quantum Hall plateaus 
carry net skymrion charge \cite{kane-90,sondhi-93,sondhi-99,fertig-94,girvin-98}.  
Thus, solutions for the ordinary ferromagnet describing 
collections of localized skyrmions carrying net charge as shown in the left
of Figs. \ref{chap:skyrmion:fig:skyrmion_single} and \ref{chap:skyrmion:fig:skyrmion_lattice} have been 
used extensively in this regime.

However, we showed in Sec. \ref{chap:skyrmion:sec:exact} that these
solutions of localized topological objects can be derived in a unified
framework along with extended stripe solutions.
There have been a number of studies on the possibility of quantum Hall states 
with stripe order.
At high Landau levels and with frozen spin degrees of freedom,
Coloumb interaction may directly favor charge density waves
as predicted theoretically \cite{koulakov-96,moessner-96} and
verified experimentally \cite{lilly-99,du-99}.  Such states
are not directly comparable to the stripe solutions we describe which
have fixed \textit{total density} and stripe order in the 
\textit{relative density}.  However, stripe order has also been proposed
\cite{demler-02,cote-02,brey-00} and experimental evidence observed
\cite{gusev-07}
in the context of quantum Hall bilayers.  Here, even though the total
density between layers is fixed, both interlayer coherence and relative
density imbalance can develop.  The isospin degree of freedom that arises
can be used to define an appropriate magnetization vector $\hat{n}$.
Here, the phase of the interlayer coherence gives the orientation
of $\hat{n}_x$, $\hat{n}_y$, while the relative density imbalance gives
$\hat{n}_z$.  States with skyrmion stripe order and winding $\hat{n}_x$,
$\hat{n}_y$ have been proposed that are direct analogs of the configurations
shown in the top row of Fig. \ref{chap:skyrmion:fig:stripes}.

For spinor condensate ferromagnets, experiments at Berkeley
suggest the possibility of a condensate with crystalline magnetic 
order \cite{dsk-08,dsk-09}.
This crystalline order arises from an effective dipolar interactions
modified by rapid Larmor precession and reduced dimensionality.
It can drive dynamical instabilities of the uniform state which occur 
in a characteristic pattern \cite{cherng-09,ueda-09}.  
Modes controlling the component $\hat{n}$
parallel (perpendicular) to the magnetic field in spin space are unstable 
along wavevectors perpendicular (parallel) to the magnetic field in real space.
Instabilities of this type can give rise to the spin textures
shown in the stripe solutions of the bottom row in Fig. \ref{chap:skyrmion:fig:stripes}.
Here, $\hat{n}_z$ is modulated along the $y$ direction while 
$\hat{n}_x$, $\hat{n}_y$ wind along the $x$ direction.

In the companion paper \cite{cherng-10b}, we have performed a systematic numerical study
of minimal energy configurations for spinor condensate ferromagnets
with dipolar interactions.  This is made possible by the use of symmetry operations
combining real space and spin space operations to distinguish different 
symmetry classes of solutions.  For applied magnetic field in the plane 
$\hat{B}=\hat{x}$ corresponding to current experiments, we show the
lowest energy configuration in Fig. \ref{chap:skyrmion:fig:dipolar}.

Notice $\hat{n}_z$ is modulated between $\pm1$ just as in the monotonic
$M$ solutions shown in Fig. \ref{chap:skyrmion:fig:stripes}.  In addition,
the $\hat{n}_x$, $\hat{n}_y$ components wind along the horizontal axis.
However, notice the winding in the $\hat{n}_x$, $\hat{n}_y$ components is not
uniform as in the solutions we find in this paper.  In addition, the winding
changes from clockwise to counter-clockwise halfway along the horizontal axis.
For the solutions we find in this paper, the skyrmion density forms stripes
of opposite charge parallel to the horizontal axis.  
For the lowest energy configuration in 
Fig. \ref{chap:skyrmion:fig:dipolar}, the non-uniform winding leads to concentration of
the skyrmion density in smaller regions and modulation in the sign
of the skyrmion density along the $x$ axis.
We find minimal energy configurations in other symmetry classes
are generally of this type with $\hat{n}_z$ oscillating
between $\pm1$ along $y$
and $\hat{n}_x$, $\hat{n}_y$ winding along $x$.  However, 
the detailed form of the winding along $x$ varies for 
different classes.

Thus we see that the exact solutions for spinor condensate ferromagnets
without dipolar interactions provides a more transparent physical picture
for the numerical solutions with dipolar interactions.  This can be seen as follows.
The solutions we find in this paper describe overall neutral collections of 
skyrmion and anti-skyrmion topological objects.  The neutrality constraint
comes from the long-ranged divergence of the skyrmion interaction and remains
even when considering additional spin interactions such as the dipolar interaction.
Morever, skyrmions and anti-skyrmions themselves have a non-trivial spin texture
which is evident in Fig. \ref{chap:skyrmion:fig:stripes} showing the solutions of this paper.
When dipolar interactions are included, such spin textures can take advantage
of the gain in dipolar interaction energy without chaning their qualitative
structure.  However, quantitative details for minimal energy configurations
such as the one shown in Fig. \ref{chap:skyrmion:fig:dipolar} require detailed
analysis of the competition between dipolar interactions, skyrmion interactions,
and spin stiffness.

In conclusion, we have presented the low-energy effective theory of spinor
condensate ferromagnets.  This effective theory describes the superfluid velocity
and magnetization degrees of freedom and can be written as a non-linear
sigma model with long-ranged interactions between skyrmions, the topological
objects of the theory.  Quantum Hall ferromagnets share a similar effective
theory with long-ranged skyrmion interactions.  For the case
of spinor condensate ferromagnets, we find exact solutions for the non-linear
equations of motion describing neutral configurations of skyrmions and anti-
skyrmions carrying zero net skyrmion charge.  These solutions describe within
a unified framework both collections of localized topological objects
as well as extended stripe configurations.  In particular, they can be used
to understand aspects of non-trivial spin textures in both quantum Hall 
ferromagnets as well as spinor condensate ferromagnets with dipolar interactions.
\begin{acknowledgements}
We thank 
D. Stamper-Kurn, M. Vengalattore, G. Shlyapnikov, S. Girvin, T.-L. Ho,
A. Lamacraft,  and M. Ueda for stimulating discussions. 
This work was supported by a NSF Graduate Research
Fellowship, 
NSF grant DMR-07-05472,  AFOSR Quantum Simulation MURI, AFOSR MURI on
Ultracold Molecules, DARPA OLE program, and Harvard-MIT CUA.
\end{acknowledgements}

\bibliography{refs}
\bibliographystyle{apsrev}

\end{document}